\documentclass[12pt]{iopart}
\usepackage[utf8]{inputenc}
\usepackage{times}
\expandafter\let\csname equation*\endcsname\relax
\expandafter\let\csname endequation*\endcsname\relax
\usepackage{amsmath}
\usepackage{amsthm}
\usepackage{amssymb}
\usepackage{amscd}
\usepackage{faktor}
\usepackage{graphicx}
\graphicspath{} 
\usepackage{epstopdf}
\usepackage{caption}
\usepackage{subcaption}
\usepackage{cancel}
\usepackage{color}
\usepackage{url}
\usepackage{appendix}
\usepackage{tikz}
\usepackage{amsfonts}
\usepackage{dsfont}
\usepackage{float}
\usepackage{comment}
\usepackage{enumerate}
\usepackage{bbm}
\usepackage{ulem}
\newtheorem{theorem}{\textsc{Theorem}}

\newcommand{\R}{\mathbb{R}}
\newcommand{\bx}{\mathbf{x}}
\newcommand{\bxp}{\bx'}
\newcommand{\posi}{\bx_{i}}
\newcommand{\posj}{\bx_{j}}
\newcommand{\bv}{\mathbf{v}}
\newcommand{\vi}{\bv_{i}}
\newcommand{\vj}{\bv_{j}}
\newcommand{\bp}{\mathbf{p}}
\newcommand{\bom}{\boldmath{\omega}}
\newcommand{\posirel}{\widetilde{\bx}_{i}}
\newcommand{\posjrel}{\widetilde{\bx}_{j}}
\newcommand{\xbar}{\overline{\bx}}
\newcommand{\rij}{\mathbf{r}_{ij}}

\newcommand{\Ni}{\mathcal{N}_{i}}
\newcommand{\st}{\, | \,}

\renewcommand{\leq}{\leqslant}

\newcommand{\rev}[1]{#1} 
\newcommand{\anonymous}[1]{#1} 
\newcommand{\mysout}[1]{} 
\providecommand{\keywords}[1]{\textbf{\textit{Key Words:}} #1}

\begin{document}
\title{The Emergence of \rev{Lines of} Hierarchy  in  Collective Motion \rev{of Biological Systems}}

\anonymous{
\author{James M. Greene$^1$, Eitan Tadmor$^2$, Ming Zhong$^3$}
\address{$^1$ Department of Mathematics, Clarkson University, Potsdam, NY, United States}
\address{$^2$ Department of Mathematics and Institute for Physical Science \& Technology, University of Maryland, College Park, MD, United States}
\address{$^3$ Department of Applied Mathematics, Illinois Institute of Technology, Chicago, IL, United States}
\ead{tadmor@umd.edu}
}
\begin{abstract}
The emergence of large scale structures in biological systems, and in particular \rev{\mysout{--}the} formation of \rev{\mysout{social} lines of} hierarchy, is observed in many scales, from collections of cells to groups of insects to herds of animals.  Motivated by phenomena in chemotaxis and phototaxis, we present a new class of alignment models which exhibit alignment into lines. The spontaneous formation  of such ``fingers"  can be interpreted as the emergence of leaders and followers in a system of identically interacting agents.  Various numerical examples are provided, which demonstrate emergent  behaviors similar to the ``fingering'' phenomenon observed in some phototaxis and chemotaxis experiments; this phenomenon is generally known as a challenging pattern to capture for existing models.  The novel pairwise interactions provides a fundamental mechanism by which agents may form social hierarchy across a wide range of biological systems.
\end{abstract}

\anonymous{\ack{Research of ET was supported in part by ONR grant N00014-2112773. Research of MZ was supported in part by NSF grant CCF-AF-2225507.}}

\submitto{\PB}
\maketitle
\ioptwocol
\section{Introduction --- emergent phenomena in biological systems}
\label{sec:introduction}
Emergent phenomena in collective dynamics are observed in a wide range of biological systems and across different scales --- from cells to bacteria, from insects to fish, from humans to mammals. Accordingly, it has been a topic of scientific interest in a wide range of disciplines, including biology, ecology, physics, mathematics, and computer science~\cite{parrish1997animal}.
In this context, one is concerned with ``\rev{active particles}'' which consists of living  agents (and likewise, certain types of mechanical agents), equipped with senses and sensors, with which they probe the environment. These are responsible for small scale pairwise interactions.
The \emph{phenomenon of emergence} is observed when a crowd of agents, driven by those small scale interactions,  is self-organized into  large scale formations: ants form colonies, insects swarm, birds fly in flocks, mobile networks coordinate a rendezvous or create traffic jams,  human opinions evolve into parties and so on.
Thus, with no apparent central control or a built-in bias in the dynamics, the question arises  --- where does this unity from within come from? what is behind the seemingly  spontaneous self-organization?\newline
Let $\phi(\posi,\posi)$ denote the amplitude of pairwise interaction  of agents positioned at $\posi$ and \rev{$\posj$}.
 Recent studies of collective dynamics  identified different classes of  interaction kernels which play a decisive role in governing the different features of their emergent behavior, \cite{motsch2014heterophilious,shvydkoy2021dynamics,tadmor2021mathematics}. These include  metric kernels depending on the metric distance between agents, \cite{cucker2007emergent,cucker2007mathematics}, 
 \begin{align}
\phi(\posi,\rev{\posj})=\phi(|\posi-\posj|).
\label{eq:phi_metric}
 \end{align}
 Then there are topologically-based kernels depending on how crowded is the region  enclosed between agents positioned at $\bx_i$ and $\bx_j$, rather than their metric distance, \cite{ballerini2008,shvydkoy2020topologically},
 \begin{align}
 \phi(\posi,\rev{\posj})=\phi\{\#k\,:\, {\mathbf x}_k\in {\mathcal C}(\posi,\posj)\}.
 \label{eq:phi_top}
 \end{align} 
 Further, we distinguish between the class of long-range heavy-tailed kernels, $\displaystyle \int_{\rev{0}}^\infty k(r){\textnormal d}r=\infty$, expressed in terms of their 
 radial envelope,~\cite{ha2008,ha2009} 
 \begin{align}
 k(r):=\min \{\phi(\bx,\bxp)\, : \, |\bx-\bxp|\leq r\},
 \label{eq:sing}
 \end{align}
  and singular-headed kernels, $k(r)=r^{-\beta}, \beta>0$,~\cite{peszek2015discrete,shvydkoy2017eulerian,do2018global} vs. short range, compactly supported kernels,  \cite{vicsek1995novel},  $k(r)\lesssim {\mathds 1}_{r\leq r_0}$.\newline
Our primary interest is in self-organization which is independent of external forces/stimuli; for a mathematical analysis of the latter see e.g.~\cite{shu2020flocking}.

\subsection{Attraction, repulsion, alignment}\label{sec:alignment}

One classifies three main types of pairwise interactions which govern  the emergent phenomena observed in biological systems, namely --- attraction, repulsion and alignment, \cite{reynolds1987flocks, vicsek2012collective,marchetti2013hydrodynamics}.  
The first two main features are \emph{attraction} which acts as a cohesion  towards average position of neighboring agents,
while \emph{repulsion} steers to avoid collisions. These are familiar from particle dynamics. A typical first order attraction-repulsion dynamics reads
\begin{align}\label{eq:positions}
\dot{\bx}_i(t)=-\frac{1}{|\Ni(t)|}\sum_{j\in \Ni(t)}\phi(\bx_i,\bx_j)(\bx_i-\bx_j).
\end{align}
Here, the agent positioned at $\bx_i$ interacts with its neighbors at 
$\Ni(t):=\{j :  \phi(\bx_i(t),\bx_j(t))\neq0\}$, of size $|\Ni|$. Thus, with the  pre-factor normalization in \eqref{eq:positions},  it can be interpreted as a local  \emph{environmental averaging} of positions. Short-range vs. long-range kernels translate into local vs. global neighborhoods. Attraction and repulsion are dictated by the positive, and respectively negative, parts of $\phi_{ij}=\phi(\bx_i,\bx_j)$. The balance between attraction and repulsion 
is responsible tor the phenomenon of  \emph{aggregation}, where  crowd of agents is   self-organized   into one or more    large scale  stationary clusters with observable geometric configuration.  Different kernels $\phi(\cdot, \cdot)$ lead to a great variety of different limiting configurations. These are observed  in cell biology, with tissue formation (mediated by cell-to-cell recognition and cell adhesion) being the prototypical example~\cite{graner2017forms}; cell aggregation also plays a fundamental role in cellular differentiation~\cite{toyoda2015cell}, proliferation~\cite{zhang2010cell,bayoussef2012aggregation}, and viability~\cite{bayoussef2012aggregation, glinel2012antibacterial}.  We mention on passing the  important role aggregation plays   in cellular viability, e.g., when it is utilized in biofilms  as a survival mechanism for bacterial cells, and for cellular adhesion in chemo- and radio-resistance \cite{green1999adhesion,croix1997cell,brown2016aggregation,lavi2013role}.  Aggregates of cells also commonly coordinate their movement to collectively migrate; prominent biological processes displaying this behavior are wound healing and cancer invasion~\cite{friedl2009collective}, as well as chemotaxis and phototaxis~\cite{theveneau2012neural,varuni2017phototaxis}.  Aggregation is of course not limited to cells; thus for example, many species of insects (e.g. monarch butterflies overwintering) and animals forming complex social structures for a diverse set of evolutionary reasons~\cite{morrell2008mechanisms}.\newline
A third main  feature in emergent dynamics is driven \emph{alignment} --- the steering towards average heading of neighboring agents. A typical second-order alignment dynamics reads
\begin{align}\label{eq:align}
\dot{\bp}_i(t)=-\frac{\tau}{|\Ni(t)|}\sum_{j\in \Ni(t)} \!\!\phi(\bx_i,\bx_j)(\bp_i-\bp_j).
\end{align}
Here, $\tau>0$ is a fixed scaling parameter, and $\bp_i$ stands for the velocity of the agent positioned at $\bx_i(t)$, \cite{cucker2007emergent,cucker2007mathematics}, $\bp_i(t) \mapsto \bv_i(t):=\dot{\bx}_i(t)$, or its orientation, \cite{vicsek1995novel,couzin2002collective,couzin2003self}, $\bp_i(t)\mapsto \bom_i(t):=\bv_i(t)/|\bv_i(t)| \in {\mathbb S}^{d-1}$.
In a typical case of long-range interactions in a  crowd of $N$ agents, $|\Ni|=N$; 
one can adjust to short- and long-range interactions, replacing 
$|\Ni| \mapsto \sum_j |\phi(\bx_i,\bx_j)|$, \cite{motsch2011}. 
The alignment encoded in  \eqref{eq:align} describes environmental averaging of velocities/orientations. Alignment  may be either local or global, depending on  the  heavy-tailed scale of the interaction kernel. 
Alignment governs emergent phenomena  of  \emph{flocking} or \emph{swarming}, found in animal populations~\cite{conradt2005consensus}, in which    agents attempt to align their heading and/or speed in a large scale  coordinated movement.  Schools of fish~\cite{krause2000fish,hemelrijk2015increased,marras2015fish}, flocks of birds~\cite{cavagna2008starflag,ballerini2008empirical,bajec2009organized,ling2019local}, and herds of animals~\cite{hughey2018challenges} are some of the most well-known examples. We mention in passing the evolutionary roles played by flocking are diverse and species dependent: examples include reproductive efficiency, predation avoidance, and route learning in migration,  \cite{goodenough2017birds,mueller2013social,riters2019birds}.  Flocking can manifest itself via \emph{synchronization}, in which  pairwise interactions of agents are  coordinated in time into large scale crowd oscillations.  Well-known examples include the frequency of flashing of firefly lights~\cite{sarfati2021self}, the ``chorusing" behavior of some species of crickets~\cite{buck1988synchronous}, and the firing of neuron cells~\cite{penn2016network}.  Flocking occurs in behavioral contexts as well, with consensus building being an emergent phenomenon in opinion dynamics~\cite{ben2005opinion}.  It is realized on many different scales, from populations of cells to populations of humans~\cite{vicsek2001question}.

The full complexity of self organization observed in biological systems is realized when \rev{combining} attraction, repulsion and alignment. 
This was originally advocated in  the pioneering work of  Reynolds \cite{reynolds1987flocks} for  realistic simulation of boids -- birds like objects.  Reynolds' model remains one of the most commonly utilized methods of describing collective motion, with extensions proposed to incorporate the effect of pheromone signaling~\cite{delgado2007use}  and obstacle avoidance~\cite{braga2018collision}, as well as a motivation for development of particle swarm optimization~\cite{kennedy1995particle}.  The incorporation of social hierarchy via leadership has also been explicitly incorporated into Reynolds' rules for boids using an additional steering force which allows agents to change the course of the flock based on the agent's position with respect to the flock~\cite{hartman2006autonomous}.  We note that although most boids models are presented as discrete velocity updates rules, they typically can be translated to either deterministic or discrete second-order systems (see Section~\ref{subsec:second_order_model}).\newline
A systematic framework for  combining attraction, repulsion and alignment mechanisms,  is offered by  \emph{anticipation dynamics} induced by a radial potential $U$, and acting at the `anticipated positions', $\bx_i^\tau:=\bx_i+\tau \bv_i$, \cite{shu2021anticipation} (here we make  the simplification of long-range interactions $|\Ni|=N$),
\begin{align}
\dot{\bv}_i(t)= -\frac{1}{N}\sum_{j=1}^N \nabla_i U(|\bx_i^\tau-\bx_j^\tau|).
\label{eq:antic}
\end{align} 
Expanding at the small ``anticipated time'' $t+\tau, \ \tau \ll 1$, one finds
\begin{align}
\begin{split}
\dot{\bv}_i(t) =& -\frac{1}{N}\sum_j \phi_{ij}(\bx_i-\bx_j)\\
 &  + \frac{\tau}{N}\sum_j \Phi_{ij}(\bv_j-\bv_i).
 \end{split}
 \label{eq:antic_expand}
\end{align}
Here, attraction and repulsion are dictated by $\displaystyle \phi_{ij}:=U'(|\bx_i-\bx_j|)/|\bx_i-\bx_j|$, and   alignment is dictated by the Hessian, $\Phi_{ij}=D^2U(|\bx_i-\bx_j|)$, with a scalar leading order term  $\psi_{ij}=U''(|\bx_i-\bx_j|)$. Thus, for example, a standard U-shape potential-based anticipation  dictates a 3Zone dynamics  in three concentric regions, ranging from interior repulsion, $U'<0$, through intermediate  alignment where $U' \sim 0$ and surrounding with exterior attraction $U'>0$. Such 3Zone dynamics is encountered in many models for flocking and swarming.
For example, many species of insects exhibit swarming behavior in which their  motion  is self-organized into approximately concentric trajectories, known as milling, or vortex formation, \cite{carrillo2008double}. This enables the insects to carry out specific tasks in the  form of collective intelligence.  Examples of swarming include the marching of locust nymphs~\cite{ariel2015locust,buhl2006disorder} and lane formation and obstacle avoidance of army ants~\cite{couzin2003self}.    Milling is most commonly associated with fish populations during schooling and mating rituals~\cite{wilson2004basking,couzin2002collective}.  \rev{It also occurs in cell clusters~\cite{newRef01, newRef02}}, and also less frequently in ants during extreme conditions~\cite{schneirla1944unique}. 

Finally we note that although not a focus of the present work, understanding collective motion for biological crowds has numerous applications in the engineering sciences.  Examples include mobile sensing networks and the utilization of cooperative unmanned aerial vehicles (UAVs)~\cite{antoniou2013congestion,hajihassani2018applications,vasarhelyi2018optimized,cao2012overview,olfati2006flocking}.
 who have only recently begun to develop quantitative theories of collective motion~\cite{couzin2009collective}.  

\subsection{A new collective model for fingering}\label{sec:fingering}
Certain forms of emergent behavior can be classified as possessing degrees of social hierarchy, where individual agents conform to distinct roles.  As with all emergent behavior, hierarchy can arise across a vast range of scales, from small groups of cells (e.g. in cell migration~\cite{vedel2013migration,qin2021roles}), to colonies of insects~\cite{brian2012social}, to extraordinarily complex systems in vertebrates~\cite{clutton2016mammal}.   A well known example occurring in bacterial motion is that of \emph{fingering}, which serves as a primary motivation for our mathematical model introduced in this work.  Fingering is a motility pattern which is often observed in cell cultures, and is characterized by cellular populations, initially undergoing essentially random and independent motion, forming structured ``finger-like" protrusions from their initial homogeneous state~\cite{alexandre2015chemotaxis,chau2017emergent,ursell2013motility}.  These protrusions indicate the emergence of social hierarchy via ``leader-type" cells at the leading edge of the protrusions; the remaining cells ``follow" in the paths determined by leading cells, often in very straight lines~\cite{ursell2013motility}.  Fingering is most closely associated with populations exposed to optical gradients (phototaxis), but is also observed in wound healing, where cellular communication is determined primarily via chemical (chemotaxis) and mechanical signaling~\cite{poujade2007collective,gov2007collective,ladoux2017mechanobiology,sengupta2021principles}.  The formation of leaders/followers is also observed in other biological systems, such as in trail formation and cooperative transport in groups of ants~\cite{czaczkes2015trail,perna2012individual,couzin2003self,gordon2019ecology,feinerman2018physics} and the marching swarms of locusts as mentioned above~\cite{ariel2015locust,buhl2006disorder}.  Many biological mechanisms exist by which leader/follower hierarchy emerges, including pheromone signaling~\cite{perna2012individual}, slime formation~\cite{ursell2013motility}, and mechanical pressure~\cite{sengupta2021principles}, although many scientific questions remain~\cite{theveneau2017leaders,kozyrska2022p53}.

It is the goal of this work to present a minimal mathematical model which describes the emergence of social hierarchy of leaders and followers via pairwise interactions; for a visualization of typical simulations exhibiting line formation, see Figure~\ref{fig:demo_2nd_yt}.  Our proposed model can be understood from a simple phenomenological perspective:  rather than metric-based interaction, $\phi_{ij}=\phi(|\bx_i-\bx_j|)$, we propose  projected-based  interactions 
\begin{align}\label{eq:lines}
\phi_{ij}=\phi(\chi_{ij}\bx_j-\bx_i), \quad \chi_{ij}:=\frac{\langle \bx_i,\bx_j\rangle}{|\bx_j|^2},
\end{align}
where the agent positioned at $\bx_i$ interacts with the \emph{traces} of neighboring agents in the forward looking cone $\bx_j\in{\mathcal N}_i:=\{\beta|\bx_i|/|\bx_j| \leq \chi_{ij} \leq 1\}$.  \rev{For a geometric illustration of the projection, see Figure~\ref{fig:projection_first_order}.}
 This leads to the spontaneous formation of leaders and followers, defined with respect to relative positions in a linear aggregate.  
 Observe that the interactions in \eqref{eq:lines} are not symmetric; further, they are not Galilean invariant. Accordingly, there is a need to shift the fixed origin, and trace the dynamics relative to center of mass, $\bx_i \mapsto \bx_i- \xbar$.
\begin{figure}[htb!]
\centering
\includegraphics[width = 0.48\textwidth]{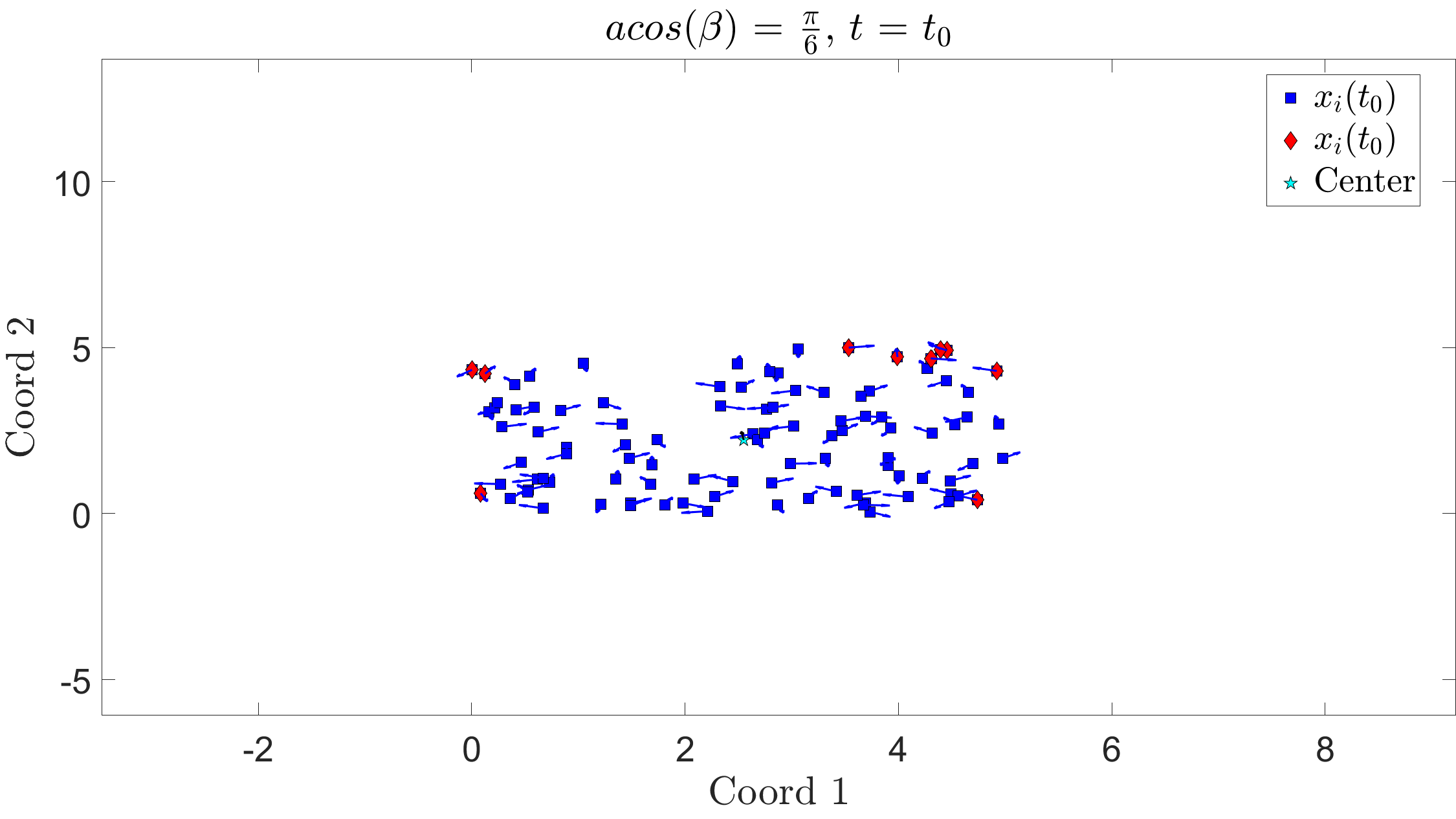}
\caption{\rev{Initial configuration of positions/velocities of second-order system as described in Section~\ref{subsec:second_order_model}.  We simulate $N=100$ agents (blue squares), with $10$ agents (red diamonds) chosen at the initial time as the furthest away from the center of mass (cyan star) at $t_0=0$.}}
\label{fig:demo_2nd_y0}
\end{figure}
\begin{figure}[htb!]
\centering
\includegraphics[width = 0.48\textwidth]{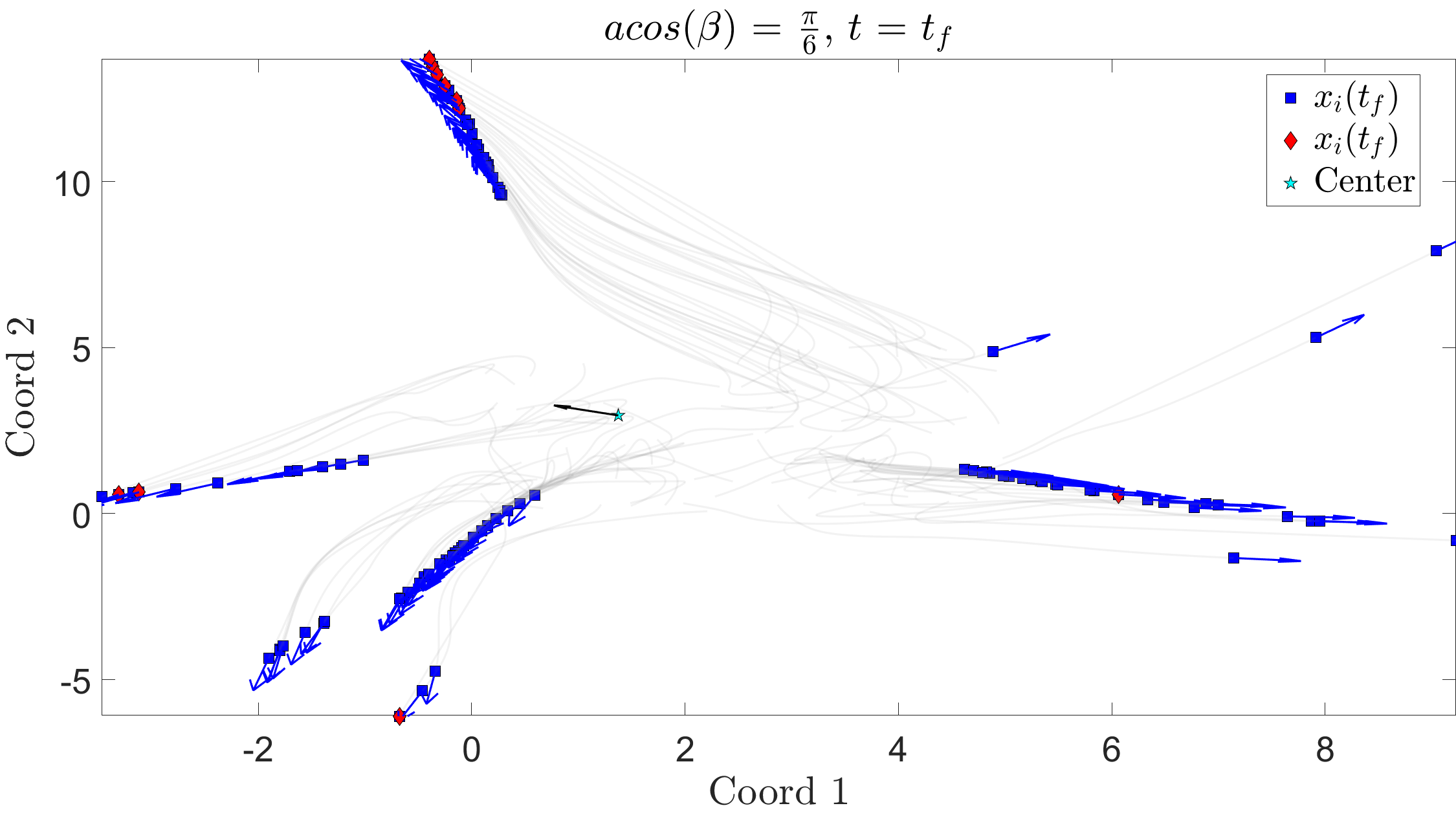}
\caption{\rev{Trajectory plot of second-order system described in Section~\ref{subsec:second_order_model} with the initial configuration shown in Figure~\ref{fig:demo_2nd_y0}.  The coloring of the agents are described in the caption of Figure~\ref{fig:demo_2nd_y0}, and the grey trailing lines indicate the path of an agent's trajectory.  Note that agents furthest away may not become ``leaders."}}
\label{fig:demo_2nd_yt}
\end{figure}
 Such interactions can be readily understood in many of the applications described above, such as the sensing of pheromone trails left by neighboring ants, and slime model deposits in bacterial cultures.  Although inspired by fingering in phototaxis and chemotaxis, the model assumes no external forcing, so that the emergence of lines is intrinsic to the interactions of the agents alone.  Furthermore, the model is sufficiently generic to describe a wide variety of phenomena, including spatial positions and velocity, but also emotions, frequencies, headings, opinions, etc. as described previously.

The remainder of the paper is organized as follows.  We provide a brief discussion of mathematical models of collective motion and chemotaxis/phototaxis in Section~\ref{sec:history}.  In Section~\ref{sec:math_model}, we provide a detailed description for the modeling framework, with details of the first- and second-order systems provided in Sections~\ref{subsec:first_order_model} and~\ref{subsec:second_order_model}, respectively.  Numerical results for each system are provided in Section~\ref{subsec:first_order_sim} and Section~\ref{subsec:second_order_sim}, and concluding remarks are provided in Section~\ref{sec:discussion_conclusions}.
\section{Alignment models of collective motion and social hierarchy}
\label{sec:history}
In this section we restrict our attention to alignment dynamics, suppressing the additional roles of attraction and repulsion. We begin with a brief overview of two alignment models; we refer to 
 \cite{vicsek2012collective,marchetti2013hydrodynamics} for a thorough discussion on the biological phenomena, and to~\cite{tadmor2021mathematics} for a recent mathematically rigorous discussion of alignment models.\newline
 The first  alignment model originates from the 1995 work of Vicsek~\cite{vicsek1995novel}, in which  self-propelled particle systems go through local averaging of  velocity orientations.  Indeed, many physical and biological systems utilize one form or another of environmental averaging~\cite{bullo2019lectures, levine2000self,d2006self,chate2008collective, chuang2007state}.\newline
A second velocity alignment model was introduced  in 2007 by Cucker and Smale~\cite{cucker2007emergent,cucker2007mathematics}. 
 The model presented in this manuscript is directly inspired by the Cucker-Smale (CS) model, so we describe it in detail here.  The system consists of $N$ identical interacting agents, each is identified  by its position $\posi$ and velocity $\vi$ in $\R^{d}$, for $i=1,2,\ldots, N$.  Their dynamics is governed by
 \begin{align}\label{eq:cs_system}
\begin{split}
    \dot{\posi}(t) &= \vi \\
    \dot{\vi}(t) &= \frac{\tau}{N}\sum_{j=1}^{N}\phi_{ij}(t)(\vj(t) - \vi(t)),
\end{split}
\end{align}
with pairwise interactions  driven by 
$\phi_{ij}(t) = \phi(\posi(t), \posj(t))$.
  The scalar \emph{communication kernel}, $\phi$,  quantifies the dynamic influence of agent $j$ on agent $i$.   In the original  CS model, the authors advocate the class of  long-range, decreasing metric kernels
\begin{align}\label{eq:phi_CS}
    \phi_{ij} &= \phi(|\posi - \posj|), \quad \phi(r)= 
                \frac{K}{(\alpha^{2}+r^{2})^{\beta}}, 
\end{align}
with constants $\ K,\beta>0$.
\rev{\mysout{We already mentioned} We previously discussed} the other classes of  singular kernels which emphasize nearby agents over those farther away~\cite{peszek2015discrete,shvydkoy2017eulerian,do2018global,minakowski2019,choi2019},
$\phi(r) = r^{-\beta}$, and the class of short-range kernels, $\phi(r)={\mathds 1}_{r \leq r_{0}}$. Metric kernels reflect, by definition, symmetric interactions, $\phi_{ij}=\phi_{ji}$, and we notice that the tacit \rev{assumption is that the communication} decays with the distance.\newline
Motivated by the original CS model, the general framework  of alignment based on pairwise interactions has inspired  considerable  work, including the hydrodynamic description of its large crowd limit, \cite{ha2008,carrillo2010asymptotic,carrillo2010particle,motsch2014heterophilious,
carrillo2014derivation,choi2017emergent}, incorporation of collision avoidance~\cite{park2010cucker}, steering~\cite{djokam2022generalized}, and stochasticity~\cite{ha2009stochastic}. 
The large-time behavior of CS alignment  dynamics \eqref{eq:cs_system} should lead the crowd to aggregate into a finite-size cluster, $\max|\bx_i(t)-\bx_j(t)|\leq D$ which in turn leads to flocking $|\bx_i(t)-\bv_j(t)|\stackrel{t\rightarrow \infty}{\longrightarrow}0$. However, left without attraction/repulsion, dynamics driven solely by alignment does not support the emergence of any preferred spatial configuration.

As mentioned in Section~\ref{sec:introduction}, the goal of this work is to provide a minimal mathematical model which exhibits the emergence of a simple form of social hierarchy through pairwise interactions.  The model is a direct analog of the CS alignment, and is inspired by the biological phenomena of fingering in chemotaxis and photoaxis.  It is advocated as a simple alignment mechanism by which \emph{a priori} identical agents evolve to form fingering structures with internal hierarchy.  It should be empathized that there is no attempt to provide our model with external environment which is of course necessary to accurately describe an externally signaled process such as phototaxis/chemotaxis; instead, we limit ourselves to cellular communication mechanisms which, we claim,  is an essential part of the  more complicated processes.  In this sense, this work is complimentary to theoretical and experimental work studying social hierarchy as well as chemotaxis/phototaxis. For example, many works formulate interacting agents systems similar to the Vicsek model~\cite{galante2011stochastic}, which may include an internal excitation variable to model phototaxis both deterministically~\cite{ha2009particle} and stochastically~\cite{bhaya2008group,levy2008modeling,levy2008stochastic}.  Slime deposition~\cite{risser2013comparative} is also a common mechanism used to describe fingering, with agent-based~\cite{varuni2017phototaxis,menon2020information} and continuum partial differential equation~\cite{ursell2013motility} proposed.  Similar approaches exist in describing chemotaxis, including modeling fingering a free boundary value problem~\cite{amar2016collective}, and extensions to the classic chemotaxis equations introduced by Keller and Segel~\cite{keller1971model,keller1971traveling,alert2022cellular}.  Hierarchy and leadership has been investigated in the CS model~\cite{shen2008cucker} as well as in network graphs with switching topologies~\cite{shao2018leader}.  Leadership arising via external signaling was introduced and analyzed in~\cite{aureli2010coordination}, \rev{moreover leadership in cells due to feedback in speed and curvature can be formed~\cite{newRef03, newRef04, newRef05},} which we note may be particularly relevant for phototaxis and chemotaxis.
\section{Mathematical models of line alignment}\label{sec:math_model}
\begin{figure}[htb!]
\centering
\includegraphics[width = 0.48\textwidth]{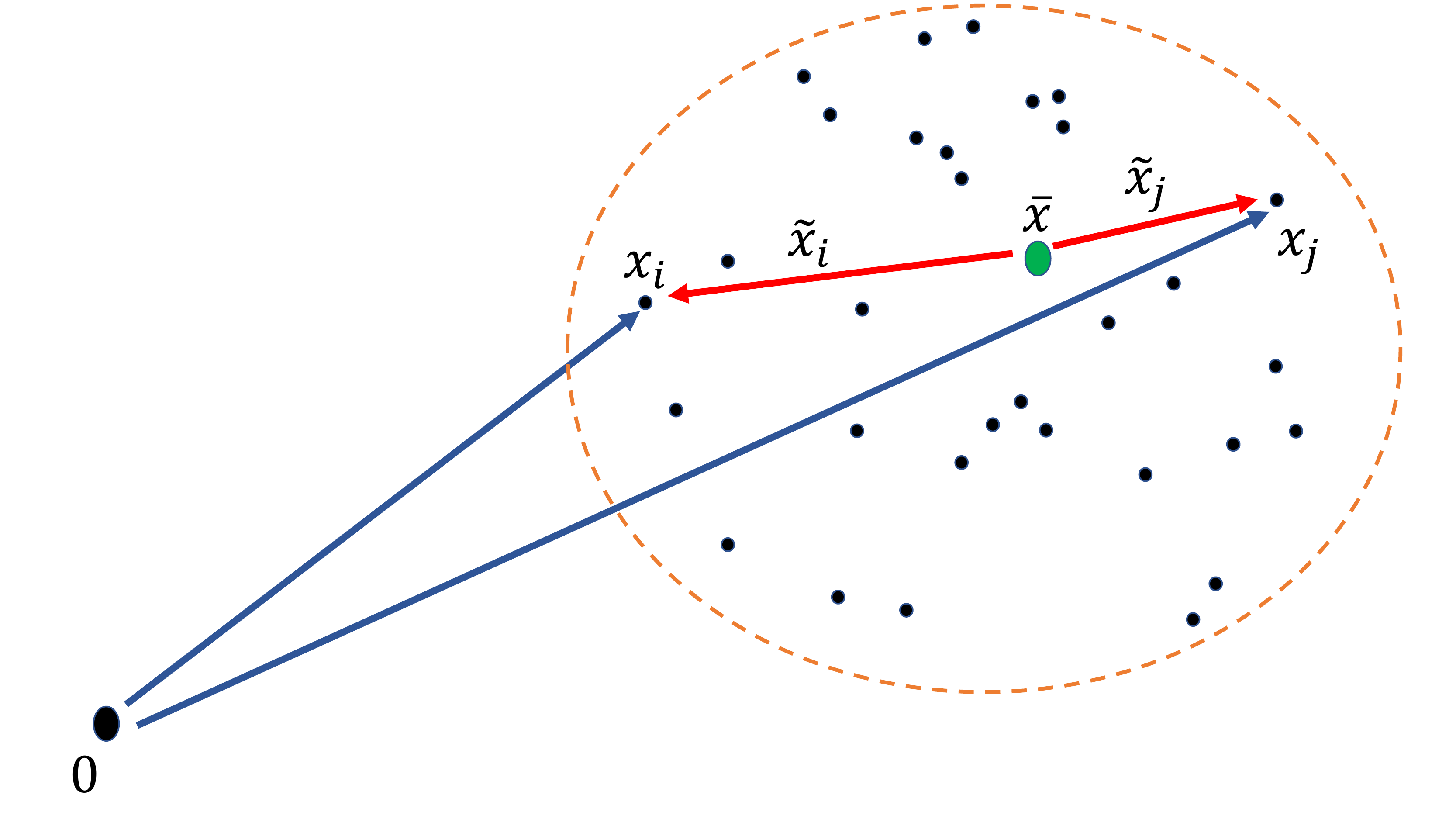}
\caption{Absolute versus relative position coordinates for agents $i$ and $j$.  Agents are indicated in black, the center of mass of the system $\xbar$ is green, absolute positions are blue, and relative positions (with respect to the center of mass) are red.  Positions vectors are indicated for two representative agents only. Note that the center of mass is calculated for the entire $N$-agent system (see~\eqref{eq:COM}).}
\label{fig:center_of_mass}
\end{figure}

Motivated by the discussion in Section~\ref{sec:introduction}, we propose both first- and second-order models which describe the emergence of hierarchical structure in interacting agent systems for active \rev{particles}, which we term generally as ``line alignment models."  For both systems, we consider a total of $N$ interacting agents.  Each agent is assigned a position  $\bx_i \in \R^d$, and,  in the case of second order models, agents  are assigned with additional   velocity,  $\bv_i \in \R^d$.  We utilize the projected position, $\chi_{ij}\bx_j$, as a way to realize the tendency of agents `to look ahead'.  In order to avoid the discussion of absolute origin, we also use the center of mass position of the whole system as the reference.  \rev{We believe this assumption is physically reasonable, as groups of bacteria/cells/animals should not utilize a global coordinate system with specified fixed origin, but rather measure positions with respect to their local environment, e.g. the center of mass of their flock, school, or other social structural unit.  Coordinate systems in local environments may be species dependent; for example, bacteria undergoing phototaxis may measure their position relative to a dominant light source~\cite{schuergers2016cyanobacteria}, while humans in a concert may measure their positions with respect to the main stage.  In an isotropic environment, a ``natural" coordinate system is the center of mass reference frame.  That is, we assume that the interacting agents measure their positions relative to the agent-system itself.}  For example, we consider the relative positions $\posirel$ and $\posjrel$ defined with respect to the center of mass $\xbar$ of the system:
\begin{align}\label{eq:COM}
    \posirel := \posi - \xbar, \quad     \xbar(t) := \frac{1}{N}\sum_{i=1}^{N}\posi(t). 
\end{align}
Here $\posi$ and $\posj$ denote the positions of the agents with respect to an arbitrary origin $0 \in \R^{d}$.  For a visualization, see Figure~\ref{fig:center_of_mass}.  \rev{We note that when interactions occur through symmetric differences of positions, as in the Cucker-Smale and Vicsek models, absolute versus relative positions result in identical dynamical systems, so that the distinction is irrelevant.  However, when considering non-symmetric interactions that arise via projected distances as in~\eqref{eq:lines}, the resulting systems possess distinct vector fields.}  Of course, certain species may indeed have global coordinate systems, such as in the mass migration of some species of birds~\cite{chernetsov2017migratory}.
\subsection{First-order model}\label{subsec:first_order_model}
We begin by introducing a first-order model, \rev{which} governs  the positions of $N$ interacting agents (cells, birds, humans, etc.).  Each agent is described by its  time-dependent position $\posi(t) \in\R^{d}$. Their dynamics is governed by pairwise interactions,
\begin{align}
    \dot{\mathbf{x}}_{i}(t) &= \frac{1}{|\Ni(t)|}\sum_{j \in \Ni(t)}\phi_{ij}(\chi_{ij}\posjrel - \posirel). \label{eq:first_order_line}
\end{align}
Here $\phi_{ij}(t)$ quantifies the interactions depending on the projected difference 
\begin{align}\label{eq:r_proj}
    \phi_{ij}:=   \phi(|\chi_{ij}\posjrel - \posirel|), \quad \chi_{ij} := \frac{\langle \posirel, \posjrel \rangle}{|\posjrel|^{2}}. 
\end{align}
 \rev{Note that in the case that $|\posjrel| = 0$ (i.e. agent $j$ is located at the center of mass of the system), the projection $\chi_{ij}$ is defined as zero.}  The neighborhood of agent positioned at $\bx_i$ is formed via a \textit{forward cone,} which models the asymmetric phenomenon of ``looking ahead",
\begin{align}
    \Ni &:= \{j  \st \beta |\posirel |   |\posjrel | \leq  \langle \posirel, \posjrel \rangle \leq |\posjrel |^{2} \}, \label{eq:Ni}
\end{align}
where  $0 < \beta \leq 1$ is a fixed constant which determines the angular size of the forward cones\footnote{For simplicity, we assume the opening of forward looking cones to be the same for all agents.}   As before, $\phi(\cdot)$ is a  \emph{communication kernel} which  quantifies the dynamic influence of the traced agent $j$ on agent positioned at $\bx_i$.   In this work, we limit ourselves to metric communication kernels.

To model line alignment, we specify both --- the pairwise interaction of agents at distance $r_{ij}=|\chi_{ij}\bx_j-\bx_i|$, as well as the spatial neighborhoods defining which agents influence the dynamics of one another.  The spatial neighborhoods are necessarily non-symmetric, but rather  ``forward-facing";  for example, if the agent positioned  at $\bx_i$ is \emph{positioned ahead} and ``in view'' of the agent positioned at $\bx_j$, then agent $j$ should be influenced by agent $i$, but not vice versa.  The notions of \emph{positioned ahead} and \emph{in view of} are quantified via $\Ni$ --- the \emph{neighborhood} of agent $i$, i.e. the set of agents which influence the dynamics of agent $i$.  

\begin{figure}[htb!]
\centering
\includegraphics[width = 0.48\textwidth]{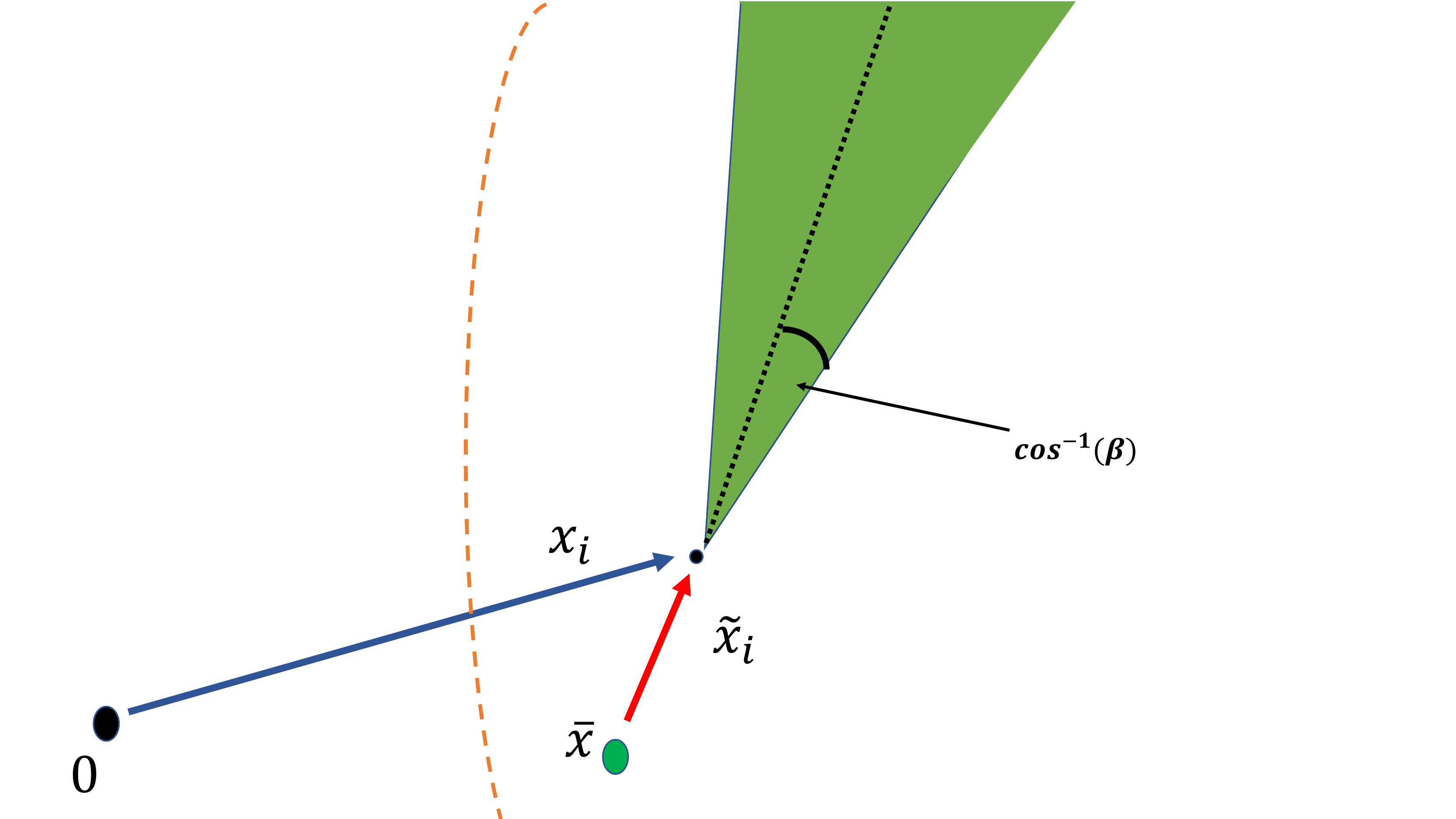}
\caption{Forward cone defining a neighborhood $\Ni$ for agent $i$ in system~\eqref{eq:first_order_line}.  The green forward conic region indicates the spatial region defining $\Ni$ in~\eqref{eq:Ni}.  Note that all positions depend on time $t$, so that conic region moves with the agents in time.}
\label{fig:neighborhood_first_order}
\end{figure}

\begin{figure}[htb!]
\centering
\includegraphics[width = 0.48\textwidth]{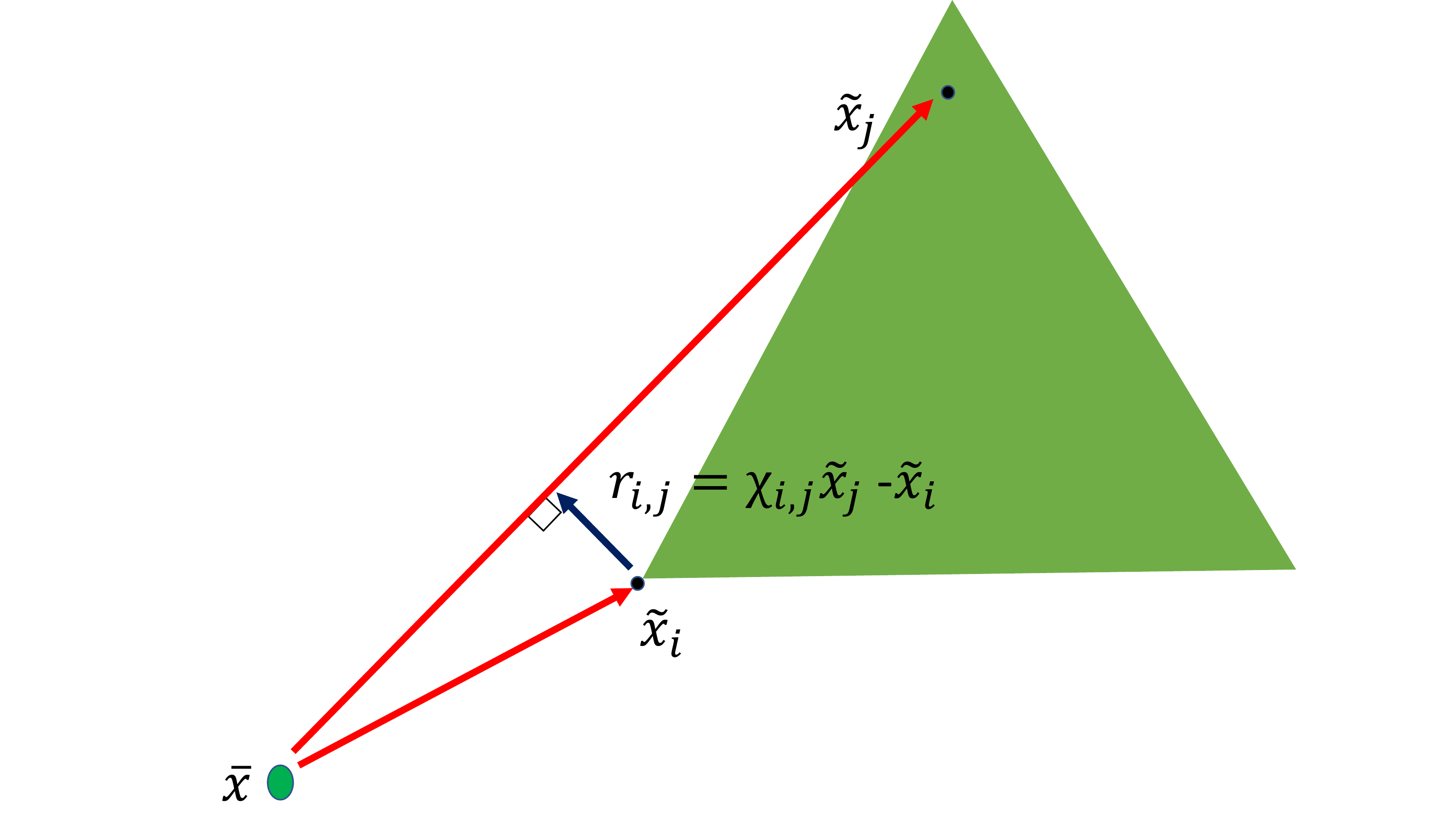}
\caption{Difference vector $\rij$ measuring the difference between $\posirel$ and its projection along $\posjrel$.  Vector $\rij$ measures how aligned agent $i$ is with respect to agent $j$.  Line alignment implies that agent $i$ acts to reduce $\rij$ for all agents $j$ in its forward cone.}
\label{fig:projection_first_order}
\end{figure}

Note that we use relative coordinates $\posirel$ and $\posjrel$ to determine the forward cone, based at the corresponding center of mass $\xbar$.  Each cone is defined via a central angle of $2\cos^{-1}(\beta)$ radians, which is symmetric about the $\posirel$ direction; this is the left-hand inequality in~\eqref{eq:Ni}.  The right-hand inequality ensures that agents are only influenced by other agents in front of them in relative position space, so that the cone is indeed forward-facing.  The latter can be understood by noting that the right-hand inequality in~\eqref{eq:Ni} restricts the length of the projection of $\posirel$ along $\posjrel$, i.e. $|\posirel| \cos(\phi) \leq |\posjrel|$ where $\phi$ is the angle between $\posirel$ and $\posjrel$, so that we require $\posjrel$ to be ahead of $\posirel$ \rev{in relation to} $\xbar$.  Since all positions $\posirel$ are time-varying, each $\Ni$  changes in time $t$; a static visualization of the conic spatial region is provided in Figure~\ref{fig:neighborhood_first_order}.
Geometrically,  the influence of agent $j$ on its dynamics, agent $i$  measures its difference in \emph{projected position} relative to agent $j$.  Thus, the form of~\eqref{eq:first_order_line} tends to align agents along \emph{lines}, as the pairwise interactions `aim' to reduce the orthogonal distance between agents $i$ and $j$.  As $\phi$ is tacitly assumed to be decreasing, agents are more  influenced by their nearer neighbors inside the forward-looking cone, i.e. with those that are more aligned with their current direction.  A visualization is provided in Figure~\ref{fig:projection_first_order}.  To completely specify the dynamics, a set of initial conditions $\{\bx_{i}(0)\}_{i=1}^{N}$ must also be prescribed.   Details on initial conditions and other parameters investigated are provided in Section~\ref{subsec:first_order_sim}.
\subsection{Second-order model}\label{subsec:second_order_model}
We can now introduce the second-order model, analogous to the first-order model detailed in Section~\ref{subsec:first_order_model}, but in which pairwise interactions influence the \textit{acceleration} of the agents; this is in contrast with~\eqref{eq:first_order_line}, where agents directly regulate velocity.  This framework is more closely related to a classical mechanics perspective, where the interaction terms are precisely interaction \textit{forces.}  The state-space is represented by the position and velocity vectors of the $N$ agents, i.e. $\{\posi, \vi\}_{i=1}^{N}$, where the second-order dynamics reads
\begin{align}
    \begin{split}
    \dot{\mathbf{x}}_{i}(t) &= \vi(t) \\
    \dot{\mathbf{v}}_{i}(t) &= \frac{1}{|\Ni(t)|}\sum_{j \in \Ni(t)}\Big( \phi(|\rij(t) |)\rij(t) \\
    &\quad + \psi(|\rij(t)|) (\vj(t) - \vi(t)) \Big)
    \end{split}
    \label{eq:second_order_line_1}
\end{align}
Here $\rij:=\chi_{ij}\posjrel-\posirel$ denotes the orthogonal component of $\posirel$ projected in the direction of $\posjrel$, so that agents interact via relative projected directions as discussed in Section~\ref{subsec:first_order_model} (see equation~\eqref{eq:r_proj}), and hence act to align ``along lines''.  The functions $\phi$ and $\psi$ characterize the strength of the interactions, both depend on the projected distances, $r_{ij}=|\rij|$, between agent currently positioned at $\posirel$, and the  agent currently positioned at $\posjrel$ which is traced to its backward  position $\chi_{ij}\posjrel$.  Interactions are again local, and the net effect on the dynamics of agent $i$ is the superposition of the interaction forces from all ``forward-looking" neighboring agents $j \in \Ni$.

The pairwise interactions  are determined by the two function $\phi$ and $\psi$.  We first observe that force governed by $\phi$ is identical to that  of the first-order attraction/repulsion in \eqref{eq:positions}.  The second term, dictated by interaction kernel $\psi$, is an extension of the Cucker-Smale velocity alignment force~\cite{cucker2007emergent, cucker2007mathematics}.    The velocity alignment in our second-order model ensures that in equilibrium, the emergent ``finger-like" lines lead to flocking, $|\bv_i(t)-\bv_j(t)|\stackrel{t\rightarrow \infty}{\longrightarrow}0$.  Indeed, the line formation will stay stable only when all the agents are moving with the same velocity; otherwise the line\rev{s \mysout{formation}formed} will not be stable. 
\section{From alignment to the emergence of lines}
\rev{The key feature of the first- and second-order models, \eqref{eq:first_order_line}, and respectively, \eqref{eq:second_order_line_1}, is the emergence of geometric structure for the trails along which the crowd is aligned --- specifically, we observe the large-time formation of curves turning into straight lines; consult the numerical simulations reported in Section ~\ref{sec:simulations} below.
A detailed analysis of this phenomenon is beyond the scope of this work and will be provided in a future work. Here we quote a prototypical result. We consider the first-order line alignment model 
\begin{align}\label{eq:line-model}
\begin{split}
\dot{\bx}_i(t)&=\frac{\tau}{\sigma_i}\sum_{j\in \Ni(t)} \!\!\!\!\phi_{ij}\big(\chi_{ij}\bx_j(t)-\bx_i(t)\big), \\
 \qquad \chi_{ij}&=\frac{\langle \bx_i,\bx_j\rangle}{|\bx_j|^2}, \quad \sigma_i:=\sum_{j\in\Ni} \phi_{ij}.
 \end{split}
\end{align}
 To simplify our discussion, we set the dynamics relative to a fixed origin, so that $\widetilde{\bx}_i \mapsto \bx_i$. The specifics of the projected-based communication, given by $\phi_{ij}=\phi(\chi_{ij}\bx_j-\bx_i)$, are not  essential; indeed it is remarkable that our results apply to a wide variety of communication protocols, independent of symmetry or occupying a  global stencil. Here, we use an adaptive normalization of  the communication protocol as in~\cite{motsch2011}, replacing $N_i \mapsto \sigma_i$, so that
$\frac{1}{\sigma_i}\sum_{j\in \Ni}\phi_{ij}=1$ (so that the dynamics does not involve  `counting' the number of agents).
\begin{theorem}\label{thm:1}
Consider the line alignment model \eqref{eq:line-model}, dictated by a decreasing kernel $\phi(r)$, which acts inside the forward-looking  cones
\[
    \Ni := \{j  \st \beta |\bx_i |   |\bx_j | \leq  \langle \bx_i \bx_j \rangle \leq |\bx_j |^{2} \}, \ \ \beta>0.
\]
Then, there exist constants, $C_0$ depending on the initial configuration and $C_\phi$ depending on $\phi$, such that the following holds, 
\[
\begin{split}
\sum_{i}\sum_{j\in \Ni}\Big(&|\bx_i(t)|^2 \cdot|\bx_j(t)|^2\\
&-|\langle \bx_i(t),\bx_j(t)\rangle|^2\Big) \leq C_0e^{-C_\phi \beta^2t}.
\end{split}
\]
\end{theorem}
Theorem \ref{thm:1} precisely quantifies the emergence phenomenon in the first-order model \eqref{eq:first_order_line}. Namely --- the crowd forms one of more distinct straight trails, led by an agent $\bx_i$ and followed by its neighbors $\{\bx_j, \ j\in\Ni\}$ so that $|\bx_i|\cdot |\bx_j|-\langle \bx_i,\bx_j\rangle \stackrel{t\rightarrow \infty}{\longrightarrow}0$. 
The  remarkable aspect of the line dynamics,  reflected in Theorem~\ref{thm:1}, is that the `kinetic energy', $\sum_{i}\sum_{j\in \Ni}|\bx_i(t)|^2 \cdot|\bx_j(t)|^2$ is easily shown to be decreasing in time. The `potential energy', however, $\sum_{i}\sum_{j\in \Ni}|\langle \bx_i(t),\bx_j(t)\rangle|^2$, does not exhibit a time-monotone behavior and may change with the configuration. It is their difference that is decreasing in time, reflecting the emergent behavior. To our knowledge, it is the first large-time, large-crowd emergence dynamics based on \emph{local} interactions  (the neighborhoods $\Ni$'s).   
}

\section{Numerical \rev{\mysout{Experiments} results}}\label{sec:simulations}
In this section, we \rev{numerically investigate the models of social hierarchy presented in Section~\ref{sec:math_model}.  Specifically, we demonstrate the emergence of lines in both the first- and second-order models, and study the effect of model parameters, including the number of agents and type of interaction kernel, on the resulting dynamics.  We also show that the formation of leader agent is indeed emergent, and cannot be easily extrapolated via initial conditions alone. \mysout{present the numerical implementation and testing of the two models presented in the previous section.  We conduct various tests on these two models to show that the emergence of lines is assured.}}
\subsection{First-order model \rev{\mysout{simulation results} line formation} }\label{subsec:first_order_sim}
\begin{figure}[htbp]
    \centering
    \includegraphics[width=0.48\textwidth]{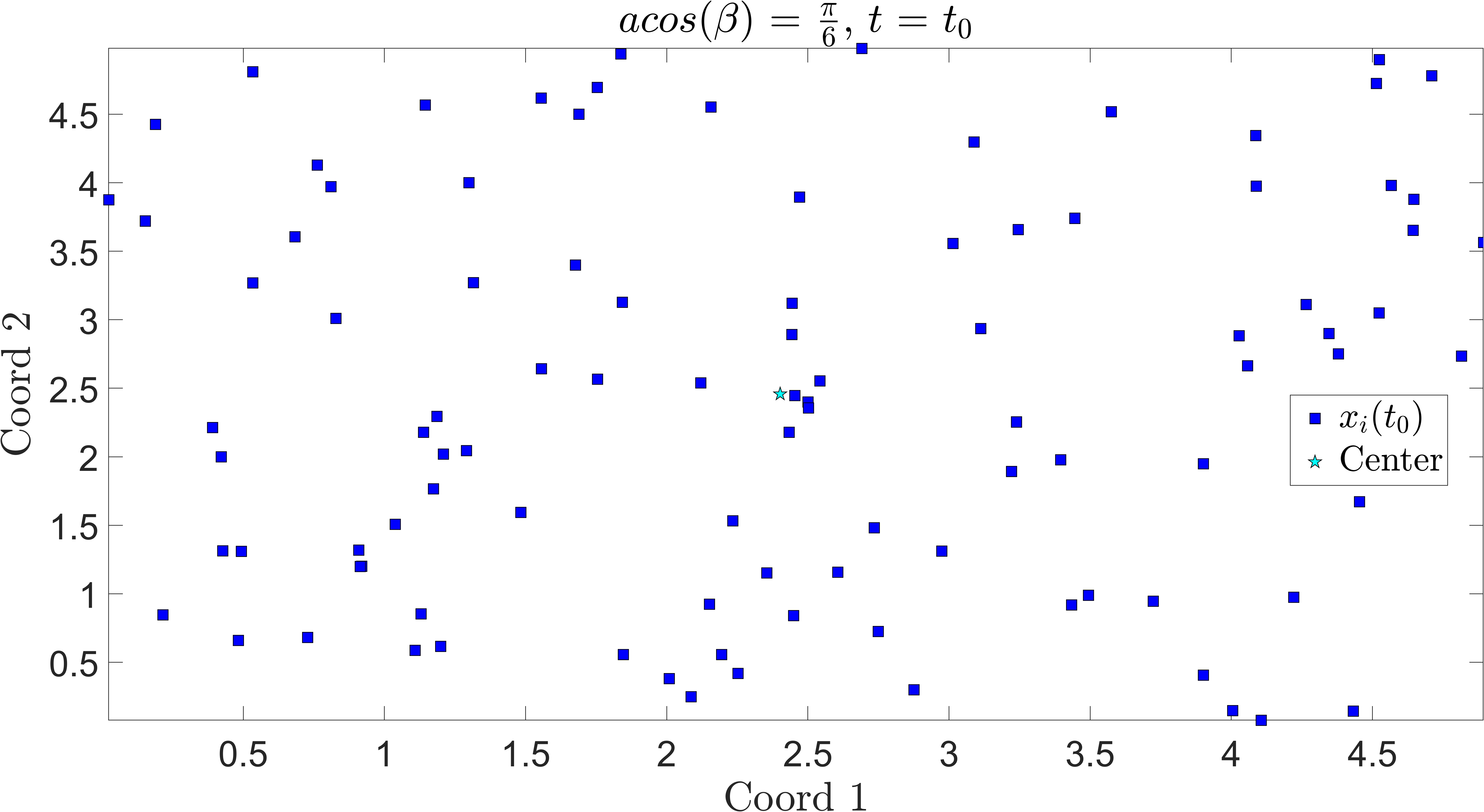}
    \caption{\rev{Initial configuration of positions for $N = 100$ agents for first-order system with parameters as in Table~\ref{table:1st_order_params}.}}
    \label{fig:result_1st_order_y0_N100}
\end{figure}

We \rev{begin by} \rev{\mysout{test}simulating} the first-order line alignment model\rev{\mysout{s}\eqref{eq:first_order_line}} with \rev{\mysout{the following}} parameters \rev{appearing in Table~\ref{table:1st_order_params}.  We are thus simulating $N=100$ agents over a period of $50$ time units in a two spatial dimensions.}  Here $\mu_0$ \rev{\mysout{is the probability distribution} denotes a probability distribution utilized} for generating the initial positions, \rev{\mysout{and} with} $\mathcal{U}([0, 5]^2)$ \rev{\mysout{is} representing} the uniform distribution over $[0, 5]^2$.  \rev{Thus, we assume that the agents are initially uniformly distributed over a square region in the plane.  A fixed realization of initial positions of agents is used for all simulations in this subsection, which is provided in Figure~\ref{fig:result_1st_order_y0_N100}. \mysout{shows one realization of the initial configuration of position for all the tests of different $\beta$.}  In this subsection, we assume a topological interaction kernel as discussed in Section~\ref{sec:introduction}:
\begin{align}
    \phi(r) &= \mathbbm{1}_{r \leq 1}(r) \quad \text{or} \label{eq:phi_first_order_sim} \\
    \phi(r) &= \frac{1}{(1 + r^2)^{0.25}}, \label{eq:phi_first_order_sim_2}
\end{align}
where $r$ denotes the projected difference between agents with respect to an agent's forward cone (see equation~\eqref{eq:r_proj}).  Recall that $\mathbbm{1}_{r \leq 1}$ denotes the indicator function on the set $[0,1]$, so that all agents with projected distance less than one unit inside of the forward cone equally influence the dynamics of the agent.  All simulations, in this and other sections, are integrated using MATLAB's built-in adaptive integrator $\text{ode}23$ for handling possible stiffness of the system.}

\begin{table}[h!]
\centering
\begin{tabular}{c | c | c | c | c | c } 
\hline
$\mu_0$                 & $d$ & $N$   & $t_0$ & $t_f$\\
\hline
$\mathcal{U}([0, 5]^2)$ & $2$ & $100$ & $0$   & $50$ \\
\hline
\end{tabular}
\caption{\rev{\mysout{Test}} Parameters \rev{utilized to simulate the first-order model as discussed in Section~\ref{subsec:first_order_sim}.\mysout{for First-order Models.}}}
\label{table:1st_order_params}
\end{table}

\rev{\mysout{We test the models for different values of $\beta$ to control the size of the cones.  The dynamics is solved using MATLAB's built-in adaptive integrator $\text{ode}23$ for handling possible stiffness of the system.  All tests use the same initial condition (shown in Figure \ref{fig:result_1st_order_y0_N100}, with the initial position $\bx_i(t_0)$ being an i.i.d sample from $\mu_0$.}}

\begin{figure}[htbp]
    \centering
    \includegraphics[width=0.48\textwidth]{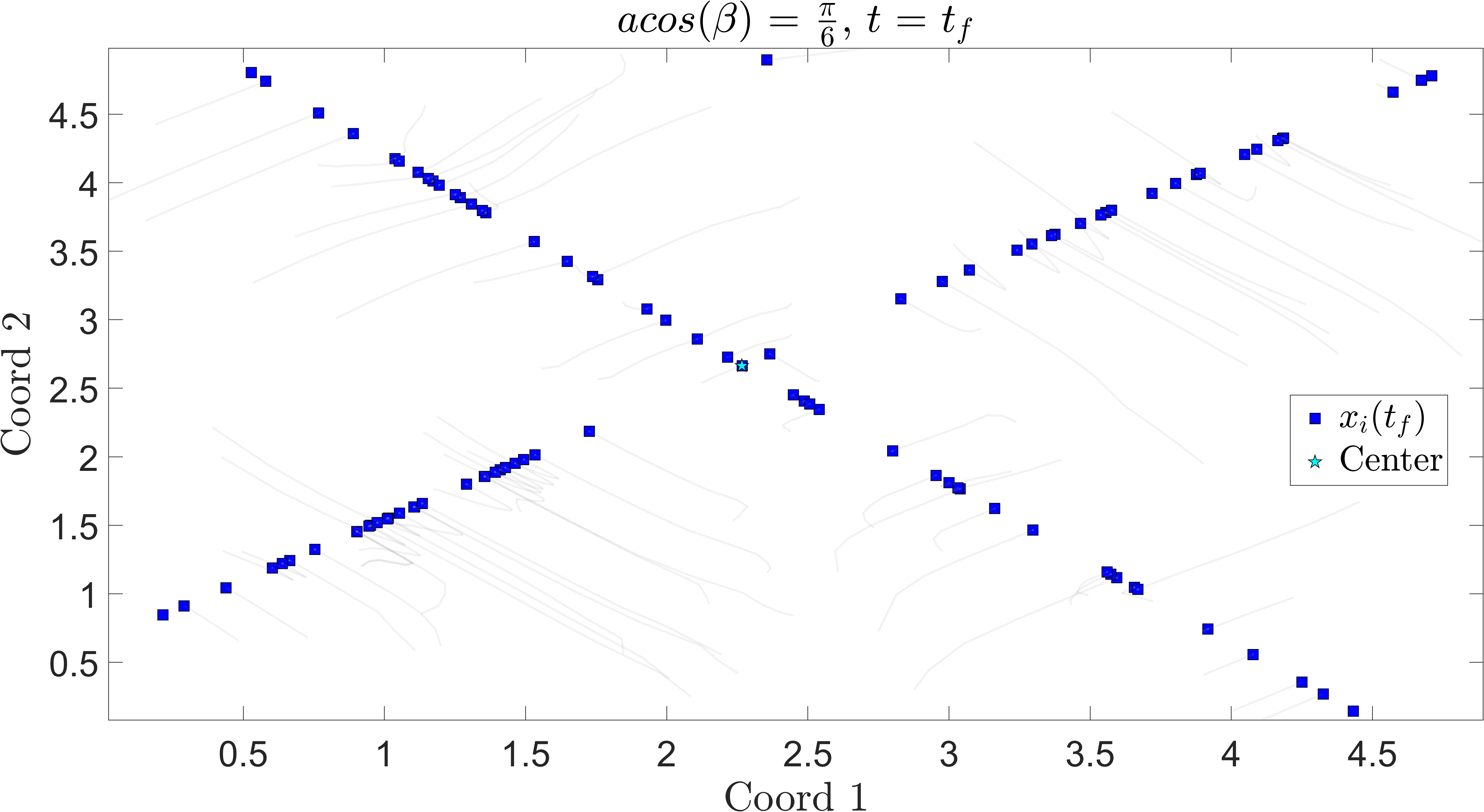}
    \caption{\rev{Trajectory plot for the first order system with parameters taken from table~\ref{table:1st_order_params} with $\beta = \frac{\pi}{6}$ and $\phi = \mathbbm{1}_{r \leq 1}$, the initial configuration of positions is shown Figure~\ref{fig:result_1st_order_y0_N100}.  We clearly observe the formation of lines.}}
    \label{fig:result_1st_order_yt_beta1_N100}
\end{figure}

\rev{We begin by demonstrating that the proposed first-order model asymptotically exhibits line formation.  Consider Figure~\ref{fig:result_1st_order_yt_beta1_N100}, which assumes a forward cone with central angle $\pi/3$ and interaction kernel given by~\eqref{eq:phi_first_order_sim}.  \mysout{The} \mysout{b}B}lue \rev{\mysout{dots} squares} in the figure\rev{\mysout{s}re}present \rev{agent positions} $\bx_i(t)$ and the \rev{\mysout{red blot} cyan star} represents the center of mass position of \rev{the \mysout{state,} system,} i.e. $\bar\bx_i(t)$.  \rev{Recall that the center of mass is not stationary, and that all agent measure relative coordinates with respect to $\bar\bx_i(t)$.  In this figure, we clearly observe the formation of spatial lines which originate from center of mass of the system.  This hierarchical structure emerges from the initial uniform distribution of positions in Figure~\ref{fig:result_1st_order_y0_N100}, and hence can be thought of as a form of emergence of social hierarchy.  Note that the ``leaders" here correspond to the agents farthest from the center of mass of the system.  The dynamics thus represent a rudimentary form of finger morphology as discussed in Section~\ref{sec:fingering}, which occurs through purely inter-agent interactions, with no reliance on external forces.  It may appear that the leader's form from those agents initially farthest from the center of mass of the system, but this is not necessarily true; see Section~\ref{subsec:leaders_emerge} for details regarding the second-order model.  We also emphasize that the system has reached equilibrium, as simulating further in time (not shown) produces the same spatial pattern observed in Figure~\ref{fig:result_1st_order_yt_beta1_N100}.}

\noindent
\rev{\mysout{\textbf{Summary}: The emergence of social hierarchy (in terms of line formation) can be clearly spotted in all of the examples of varying $\beta$ (varying sizes of the neighborhood).  The number of lines (in this case $6$ in total) is fixed for all different $\beta$ and all of the them start from the center of mass position (shown in a red dot).}}  

\subsection{\rev{Variation of dynamics as a function of forward cone size}}\label{subsec:first_order_vary_cone}

\begin{figure}[htbp]
    \centering
    \includegraphics[width=0.48\textwidth]{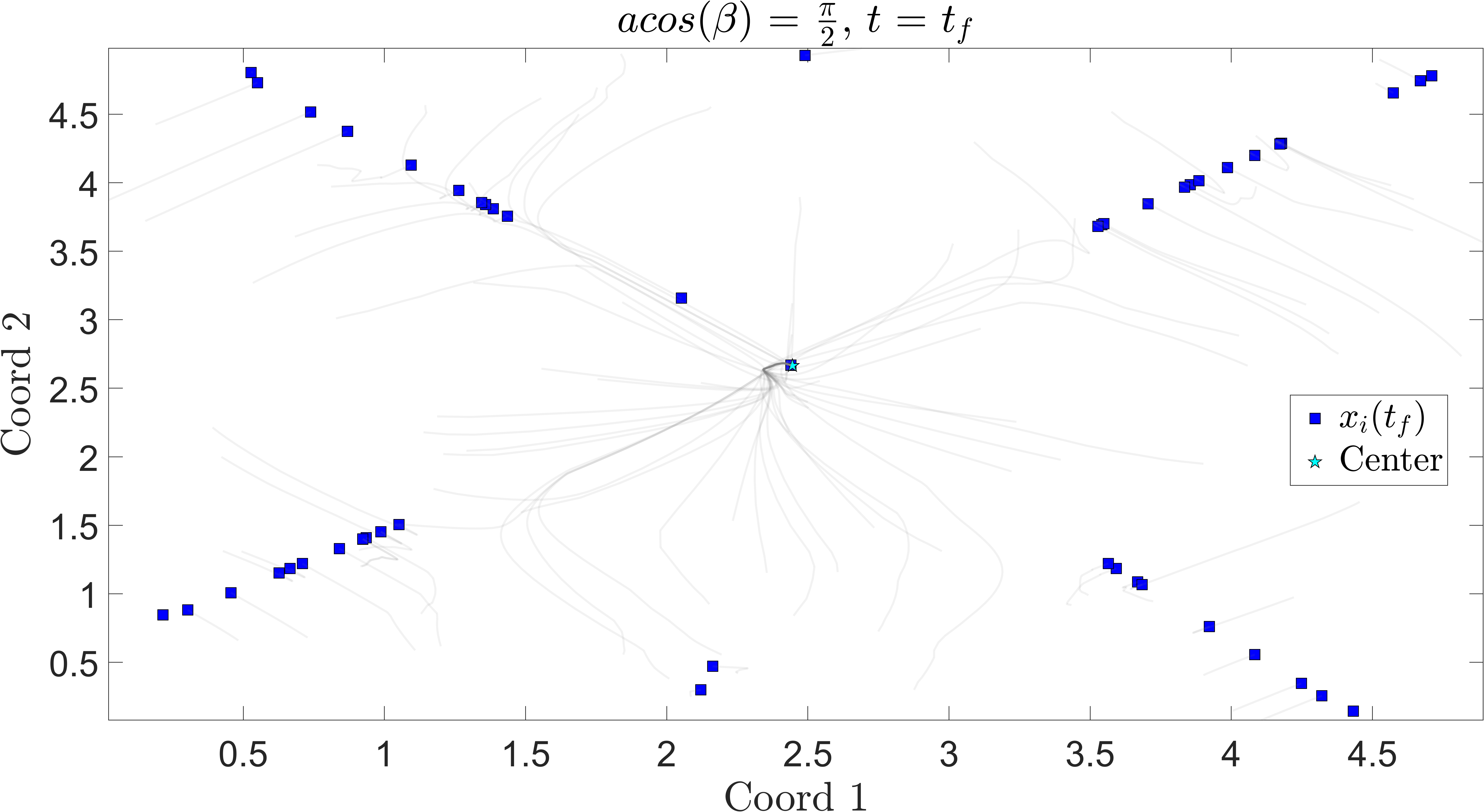}
    \caption{\rev{Trajectory plot for the first-order system with parameters taken from Table~\ref{table:1st_order_params} with $\beta = \frac{\pi}{2}$ and $\phi = \mathbbm{1}_{r \leq 1}$, and initial configuration of positions is shown in Figure~\ref{fig:result_1st_order_y0_N100}. We clearly observe the formation of lines, with significant clustering about the center of mass (compare to Figure~\ref{fig:result_1st_order_yt_beta1_N100}).}}
    \label{fig:result_1st_order_yt_beta2_N100}
\end{figure}

\rev{We investigate the dynamics of line formation as a function of the size of the forward cone.  Specifically, we are interested in understanding the effect of the angular size of the cone, defined by $\beta$ (see Figure~\ref{fig:neighborhood_first_order}), on the dynamics line formation. Questions of specific interest are both the the number and density of lines formed, which thus correspond to the number of emergent leaders (equivalently, the number of ``fingers" formed).  Intuitively, we expect that the number of lines formed should increase as the size of the forward cone decreases, as each agent acts with a higher degree of locality.  For a demonstration, see Figure~\ref{fig:result_1st_order_yt_beta2_N100}, which utilizes the same initial conditions and parameters as in Figure~\ref{fig:result_1st_order_yt_beta1_N100}, with the exception that a central angle corresponding to $\pi$ defines the forward cone (note that in the case, it is really a forward \textit{plane}, and not a cone).   We observe a similar pattern of lines compared to the smaller, and hence more local, forward cone, but with a significantly higher degree of clustering to points, so that the asymptotic behavior is more similar to \textit{points}, as opposed to lines.  The mechanisms that produces this behavior is due to a combination of two factors:  1) the dynamics are first-order, and hence tend to exhibit aggregation, and 2) the interaction kernel~\eqref{eq:phi_first_order_sim} assumes a uniformity of influence with respect to all neighboring agents.  To see the resulting pattern formation, and the effect of varying $\beta$, for the CS interaction kernel~\eqref{eq:phi_first_order_sim_2}, see Section~\ref{subsec:phi_variation}.  }

\subsection{\rev{Variation of dynamics as a function of initial conditions}}

\begin{figure}[htbp]
    \centering
    \includegraphics[width=0.48\textwidth]{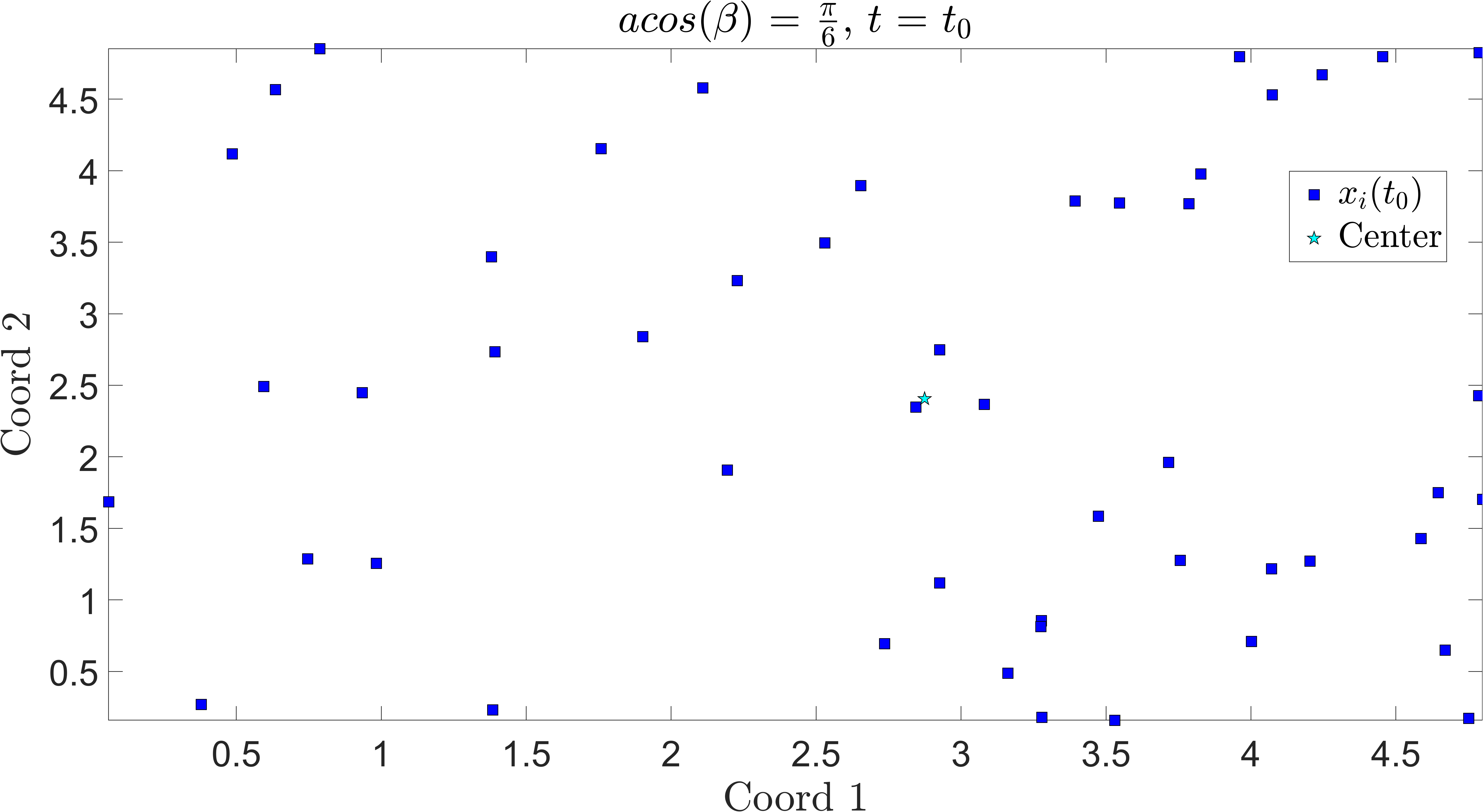}
    \caption{\rev{Initial configuration of position for first-order system with the parameters taken from Table~\ref{table:1st_order_params}, except for a smaller number of agents ($N = 50$).}}
    \label{fig:result_1st_order_y0_N50}
\end{figure}

\begin{figure}[htbp]
    \centering
    \includegraphics[width=0.48\textwidth]{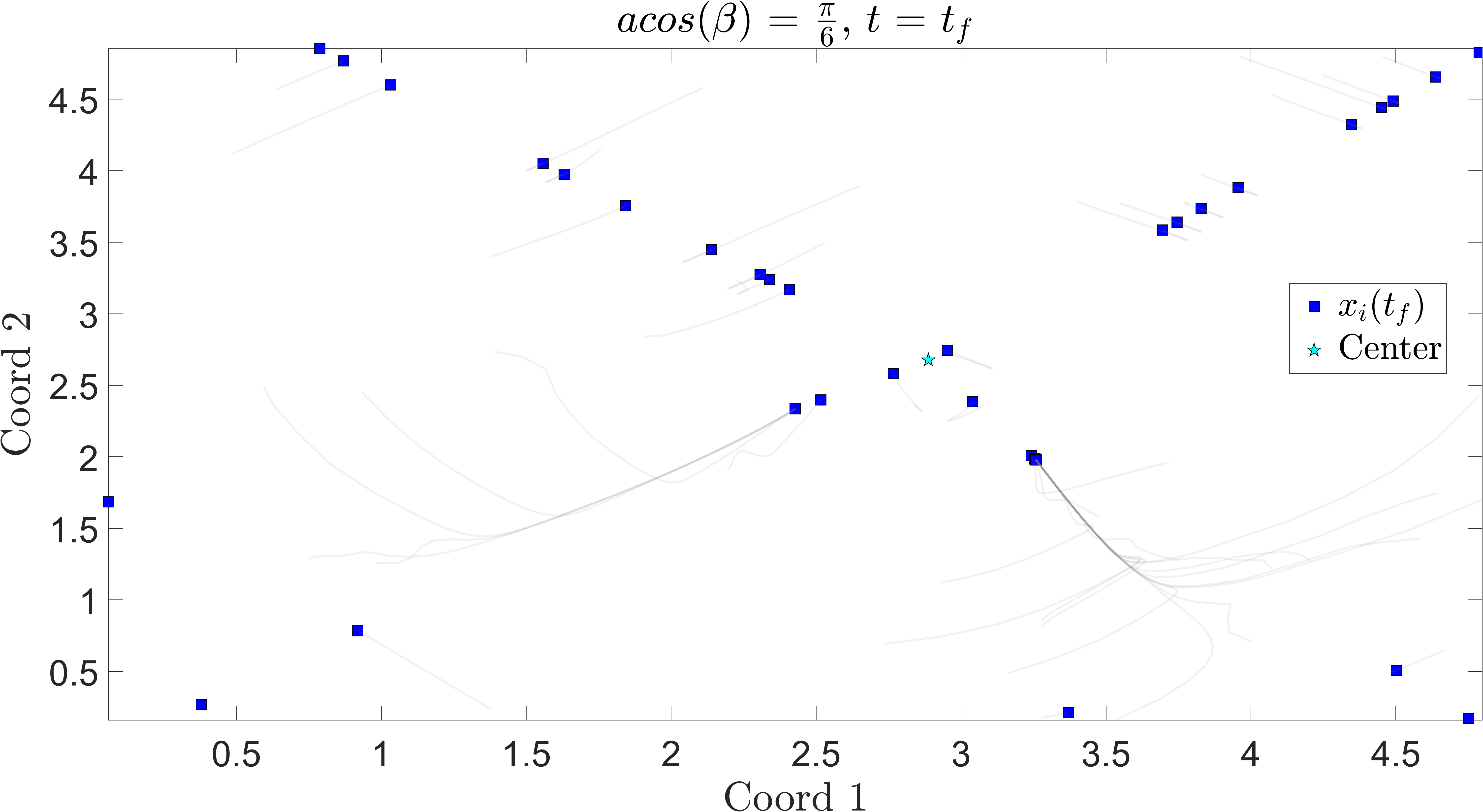}
    \caption{\rev{Trajectory plot for the first-order system with parameters taken from Table~\ref{table:1st_order_params} with $\beta = \frac{\pi}{6}$ and $\phi = \mathbbm{1}_{r \leq 1}$, and initial configuration of positions is shown in Figure~\ref{fig:result_1st_order_y0_N50}.}}
    \label{fig:result_1st_order_yt_beta1_N50}
\end{figure}

\begin{figure}[htbp]
    \centering
    \includegraphics[width=0.48\textwidth]{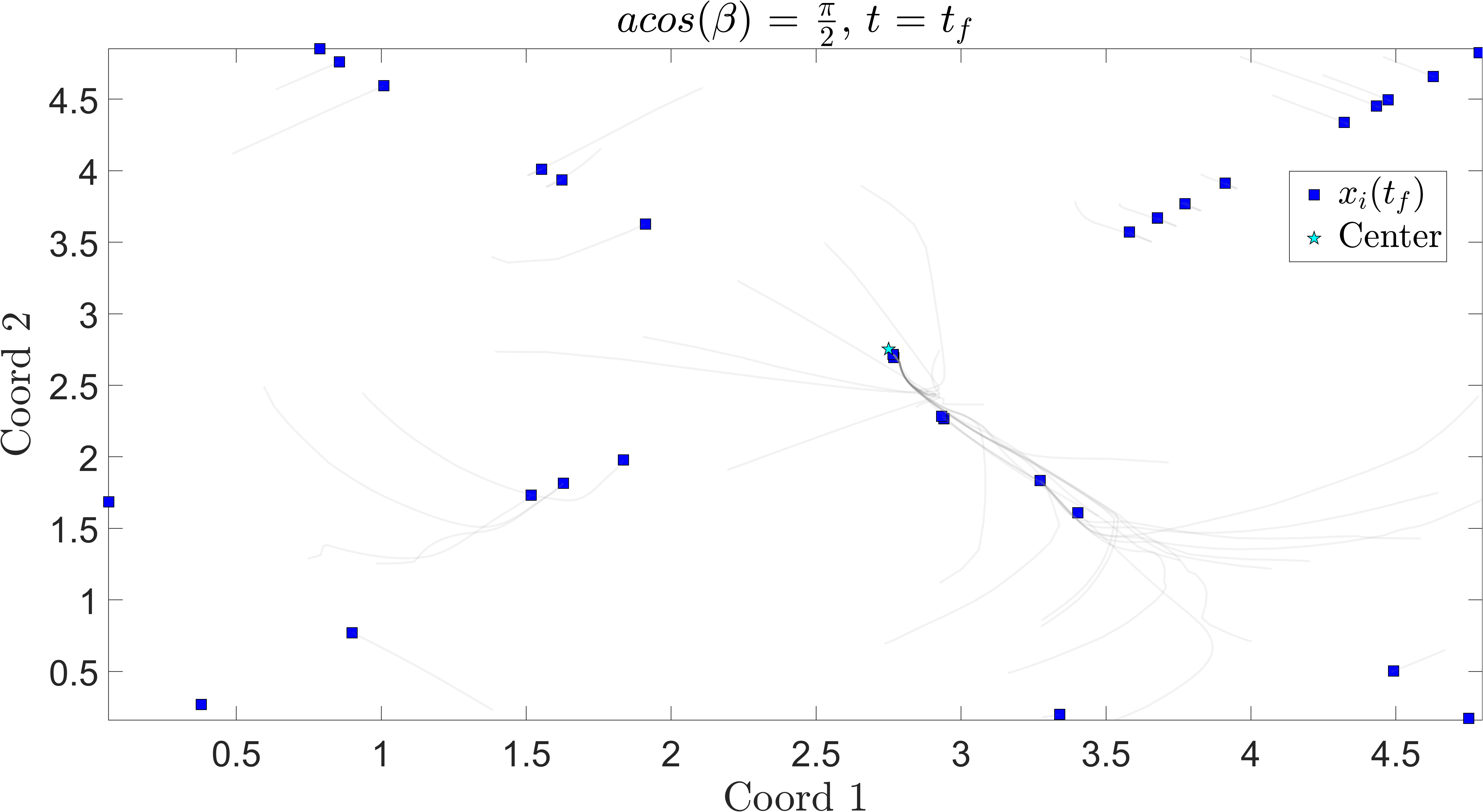}
    \caption{\rev{Trajectory plot for the first-order system with parameters taken from table~\ref{table:1st_order_params} with $\beta = \frac{\pi}{2}$ and $\phi = \mathbbm{1}_{r \leq 1}$, and initial configuration of positions is shown in Figure~\ref{fig:result_1st_order_y0_N50}.}}
    \label{fig:result_1st_order_yt_beta2_N50}
\end{figure}

\rev{In the previous section, we studied line formation for a fixed set of initial conditions.  Also of interest is the role of initial conditions on the distribution of lines.  More precisely:  is the resulting pattern robust or highly sensitive to the agent's initial positions?  Numerical simulations (not provided) suggest that the final configuration of lines (both number of lines, and orientations) are highly dependent on the initial conditions; resampling $\mu_{0}$ generally results in a different equilibrium distribution.  We note that this is not surprising, and it is common feature of models describing collective motion.  Similarly, we investigate how the behavior changes as a function of number of agents ($N$ in Table~\ref{table:1st_order_params}).  This a natural scientific question, as line formation occurs across a variety of scales; for example, the number of ants composing a trail be on the order of one hundred, while the number of bacteria generating finger morphology in phototaxis may be on the order of one thousand.  Furthermore, such questions are of mathematical interest, as they may provide insight into corresponding coarse-grained macroscopic models, such as mean-field and hydrodynamic limits, described by Boltzmann-type partial differential equations.  We thus investigate to what degree the social hierarchy model proposed in this work is dependent on the number of agents in the system.  As an example, we repeat simulations appearing in Figures~\ref{fig:result_1st_order_yt_beta1_N100} and~\ref{fig:result_1st_order_yt_beta2_N100}, with fewer agents ($N=50$ versus $N=100$ previously; new initial conditions are provided in Figure~\ref{fig:result_1st_order_y0_N50}); corresponding results can be found in Figures~\ref{fig:result_1st_order_yt_beta1_N50} and~\ref{fig:result_1st_order_yt_beta2_N50}.  Similar qualitative dynamics are apparent for the smaller system, but in general we see that the distributions of lines is quite different, even for the corresponding forward cones.  }

\subsection{\rev{Effect of interaction function on pattern formation}}
\label{subsec:phi_variation}
\begin{figure}[htbp]
    \centering
    \includegraphics[width=0.48\textwidth]{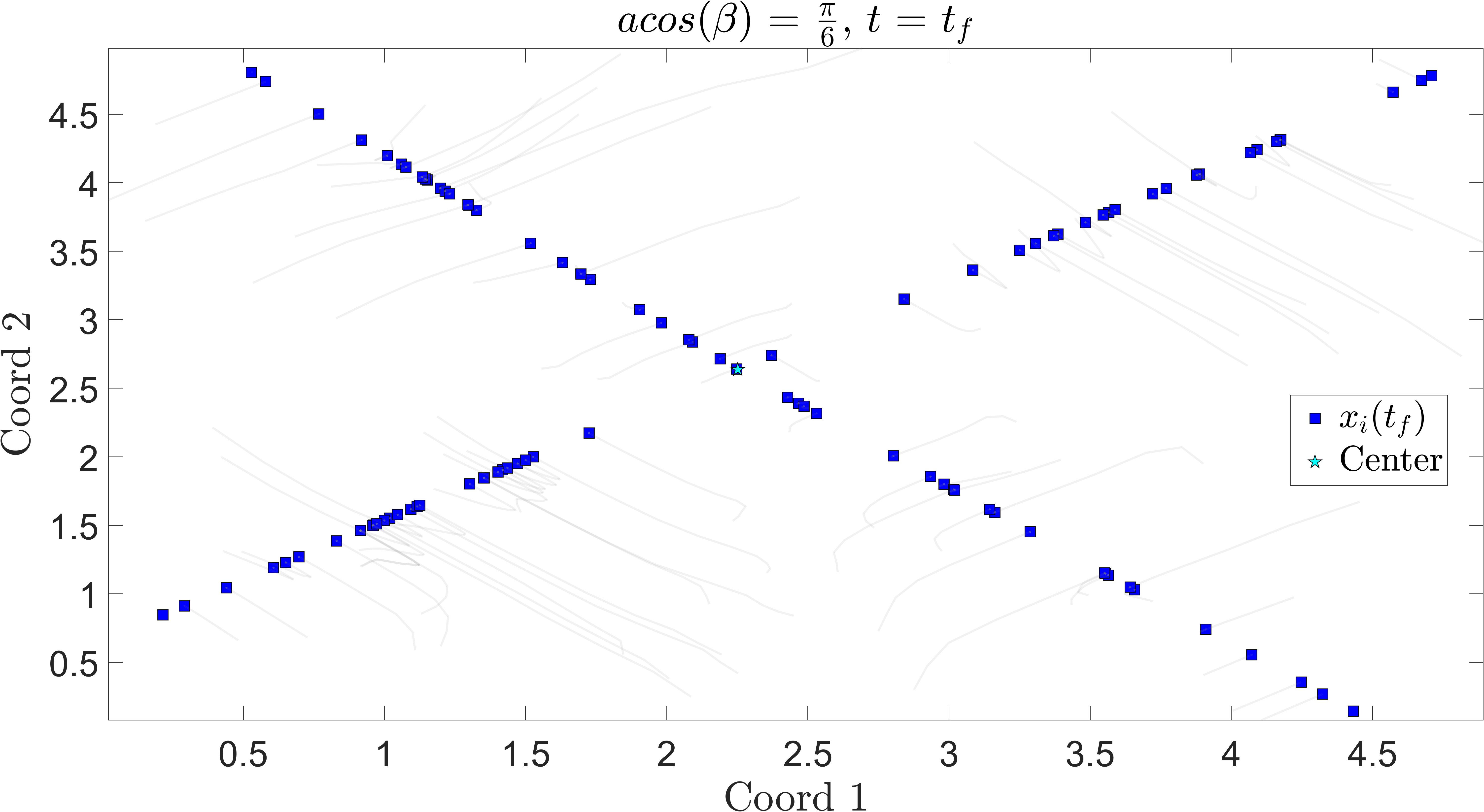}
    \caption{\rev{Trajectory plot for the first-order system with parameters taken from Table~\ref{table:1st_order_params} with $\beta = \frac{\pi}{6}$ and $\phi = \frac{1}{(1 + r^2)^{0.25}}$, and initial configuration of positions is shown in Figure~\ref{fig:result_1st_order_y0_N100}.}}
    \label{fig:result_1st_order_yt_beta1_N100_phi2}
\end{figure}

\begin{figure}[htbp]
    \centering
    \includegraphics[width=0.48\textwidth]{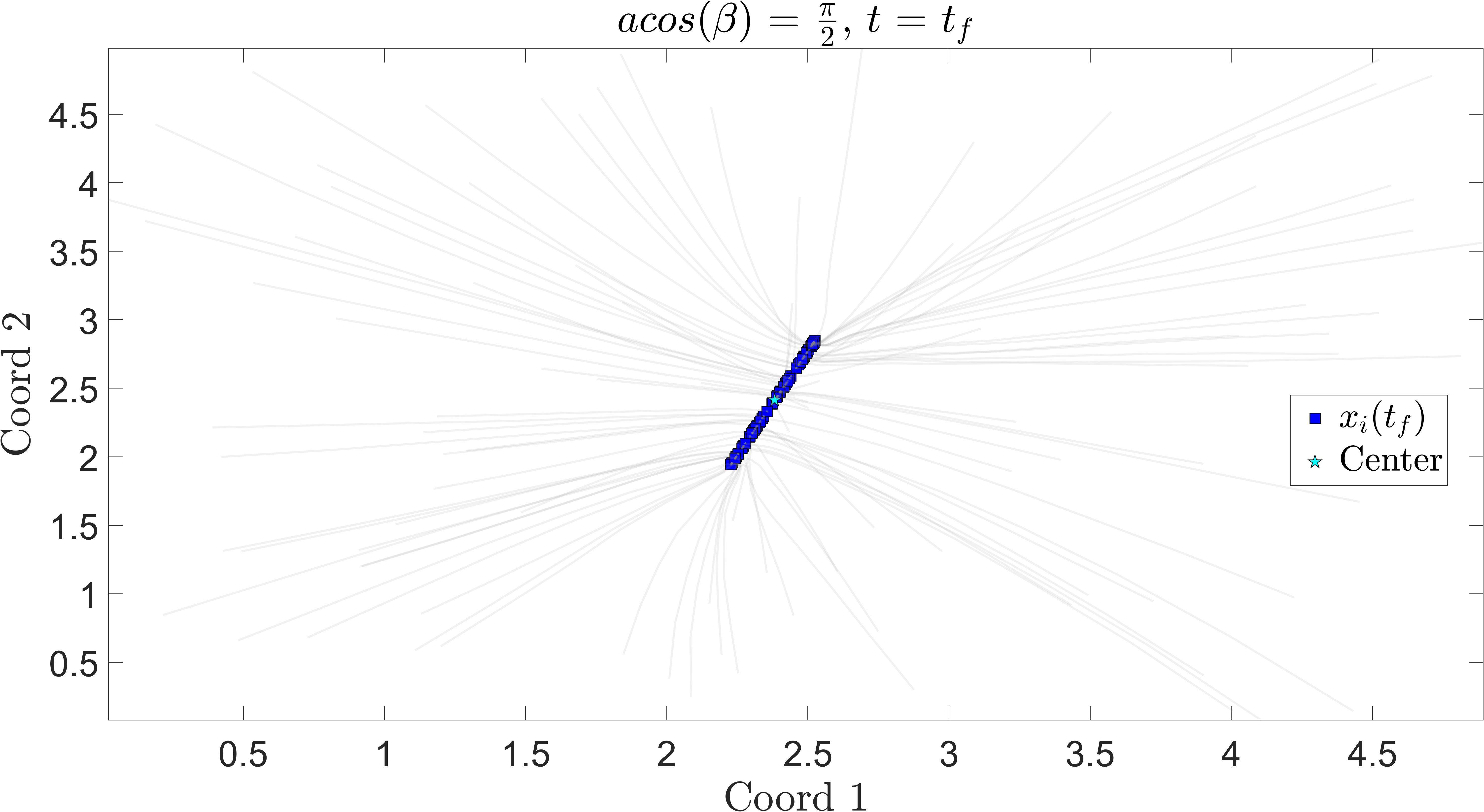}
    \caption{\rev{Trajectory plot for the first-order system with parameters taken from Table~\ref{table:1st_order_params} with $\beta = \frac{\pi}{2}$ and $\phi = \frac{1}{(1 + r^2)^{0.25}}$, and initial configuration of positions is shown in Figure~\ref{fig:result_1st_order_y0_N100}.}}
    \label{fig:result_1st_order_yt_beta2_N100_phi2}
\end{figure}

\rev{In the previous sections, we assumed a topological interaction kernel given by~\eqref{eq:phi_first_order_sim}, which weights all neighboring agents equally in a forward cone with a limited support. We here instead use a global interaction function, i.e. $\phi = \frac{1}{(1 + r^2)^{0.25}}$, to demonstrate the effect of the interaction function on line formation for two different $\beta$ values.  As discussed in Section~\ref{subsec:first_order_vary_cone}, we expect 
a global kernel to exhibit a more pronounced response to the forward cone size with respect to angle $\beta$.  Results of simulations are provided in Figures~\ref{fig:result_1st_order_yt_beta1_N100_phi2} and~\ref{fig:result_1st_order_yt_beta2_N100_phi2}.  Compared to the locally supported $\phi$ (Figures~\ref{fig:result_1st_order_yt_beta1_N100} and ~\ref{fig:result_1st_order_yt_beta2_N100}), this system's final distribution of lines shows considerable variation:  two lines for $\beta=\pi/6$ become one for $\beta=\pi/2$.  Note that the initial configuration of position is the same as shown in Figure~\ref{fig:result_1st_order_y0_N100}.}  


\subsection{Second-order model simulation results}\label{subsec:second_order_sim}
\rev{\mysout{We show one example of the second order model using the following parameters.} We next simulate the second-order model~\eqref{eq:second_order_line_1} to demonstrate the dynamics of line formation in a system where interactions affect acceleration; simulation details are provided in Table~\ref{table:2nd_order_params}.  Recall from Section~\ref{subsec:second_order_model} that we must specify two interaction kernels:  $\phi$, which governs the inter-agent interaction force of line alignment, and $\psi$, which accounts for velocity alignment and hence stabilization.  Note that $\phi$ is completely analogous to the first-order model, except it more directly corresponds to a classical Newtonian force law, as the system is second-order.  We assume the same function dependence~\eqref{eq:phi_first_order_sim} for $\phi$, and assume the classical Cucker-Smale interaction kernel for $\psi$:
\begin{align}
    \psi(r) &= \frac{1}{(1+r^{2})^{1/2}}.
    \label{eq:psi_cs_second_order}
\end{align}
} 

\begin{table}[h!]
\centering
\begin{tabular}{c | c | c | c | c | c | c | c } 
\hline
$\mu^{\bx}_0$ & $\mu^{\bv}_0$  & $d$ & $N$   & $t_0$ & $t_f$ & $\phi$ & $\psi$\\
\hline
$[0, 5]^2$    & $\mathbb{B}^2$ & $2$ & $100$ & $0$   & $10$ & $\mathbbm{1}_{r \leq 1}$ & $\frac{1}{(1+r^{2})^{1/2}}$\\
\hline
\end{tabular}
\caption{\rev{\mysout{Test}} Parameters \rev{utilized to simulate the second-order model as discussed in Section~\ref{subsec:second_order_sim}.\mysout{for Second-order Models.}}}
\label{table:2nd_order_params}
\end{table}

\begin{figure}[htbp]
    \centering
    \includegraphics[width=0.48\textwidth]{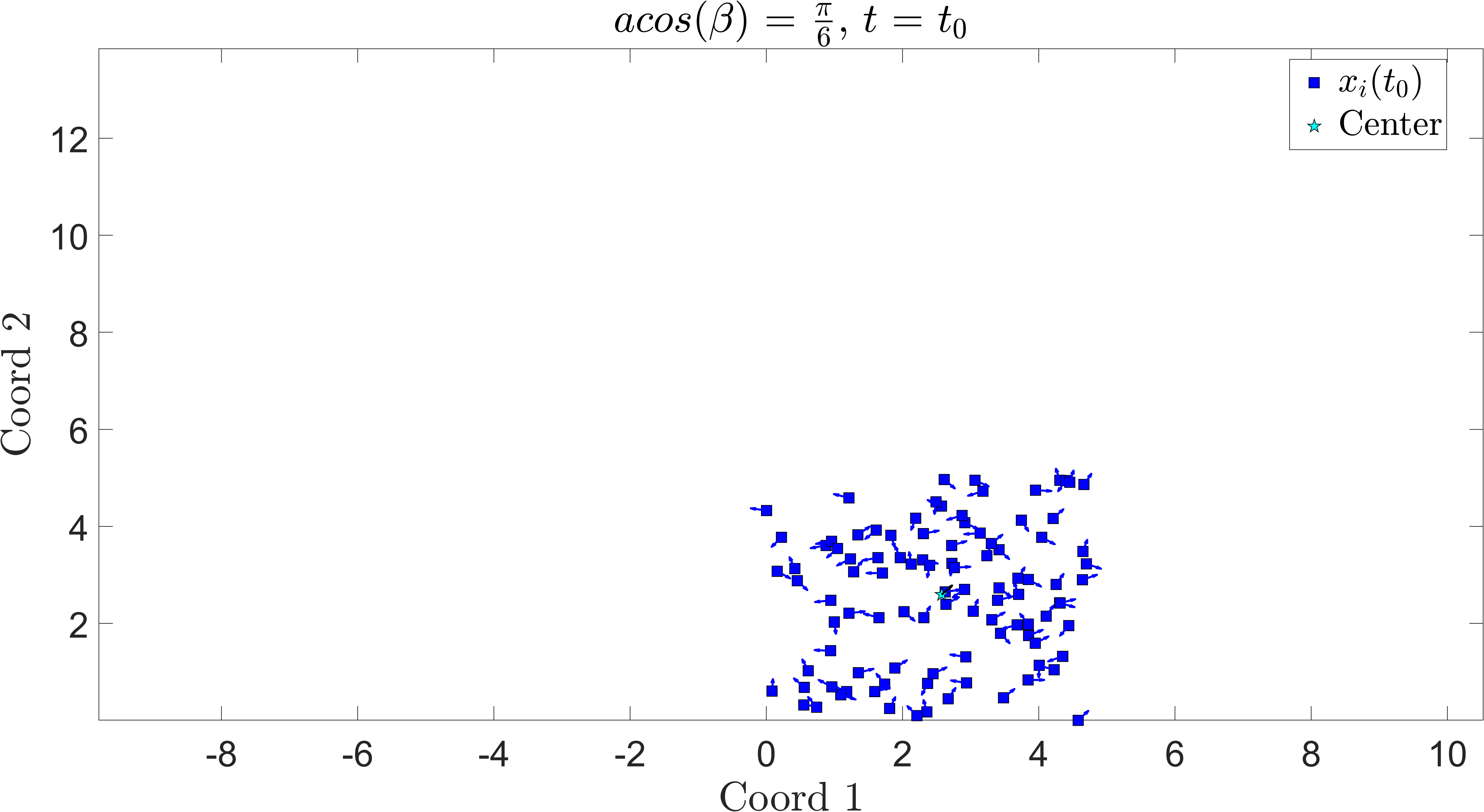}
    \caption{\rev{Initial configuration of positions and velocities for second-order system of $N = 100$ agents with the parameters taken from Table~\ref{table:2nd_order_params}.}}
    \label{fig:result_2nd_order_y0_N100}
\end{figure}

\begin{figure}[htbp]
    \centering
    \includegraphics[width=0.48\textwidth]{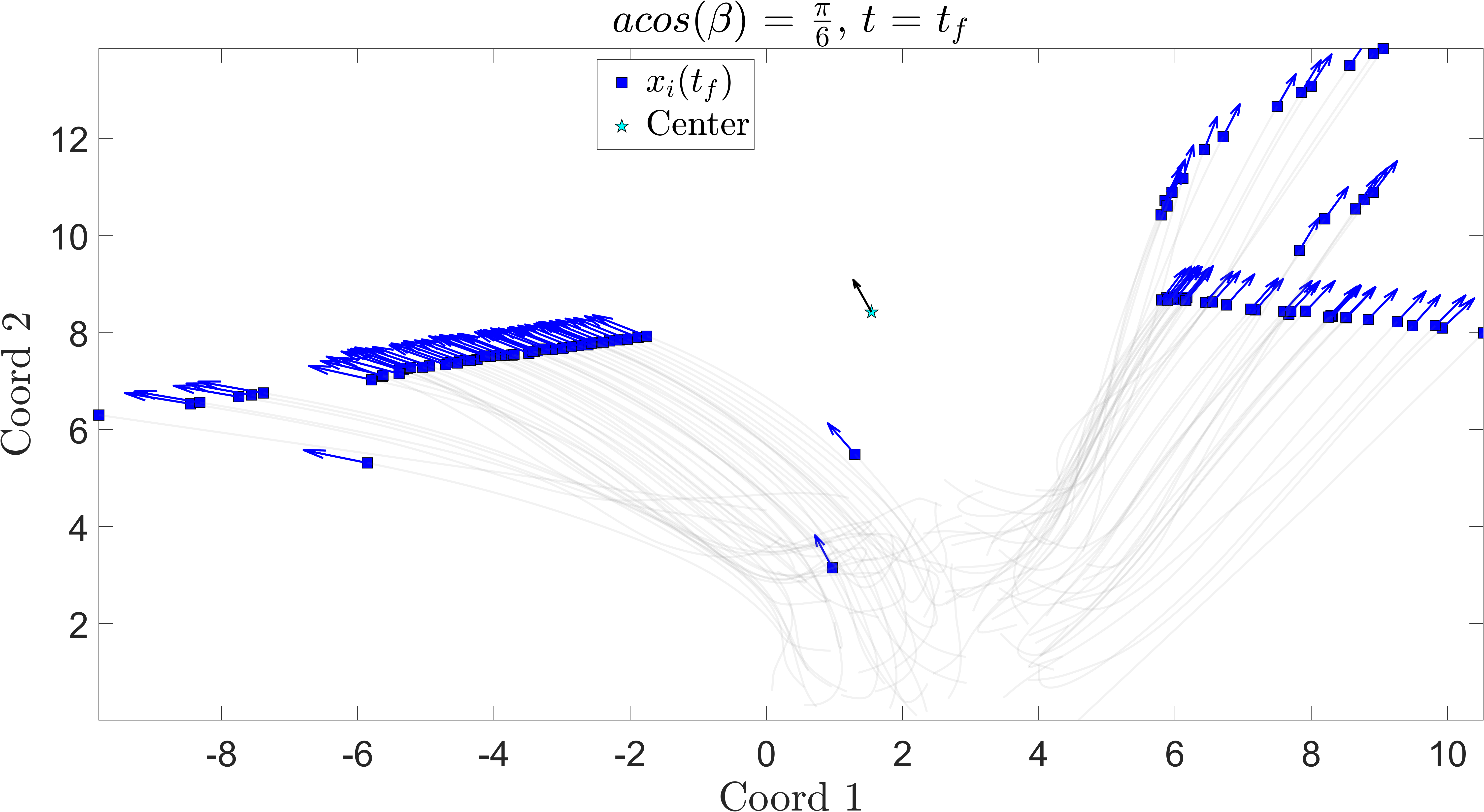}
    \caption{\rev{Trajectory plot for the second-order system with parameters taken from Table~\ref{table:2nd_order_params} with $\beta = \frac{\pi}{6}$, and initial configuration of positions/velocities is shown in Figure~\ref{fig:result_2nd_order_y0_N100}.}}
    \label{fig:result_2nd_order_yt_beta1_N100}
\end{figure}

\begin{figure}[htbp]
    \centering
    \includegraphics[width=0.48\textwidth]{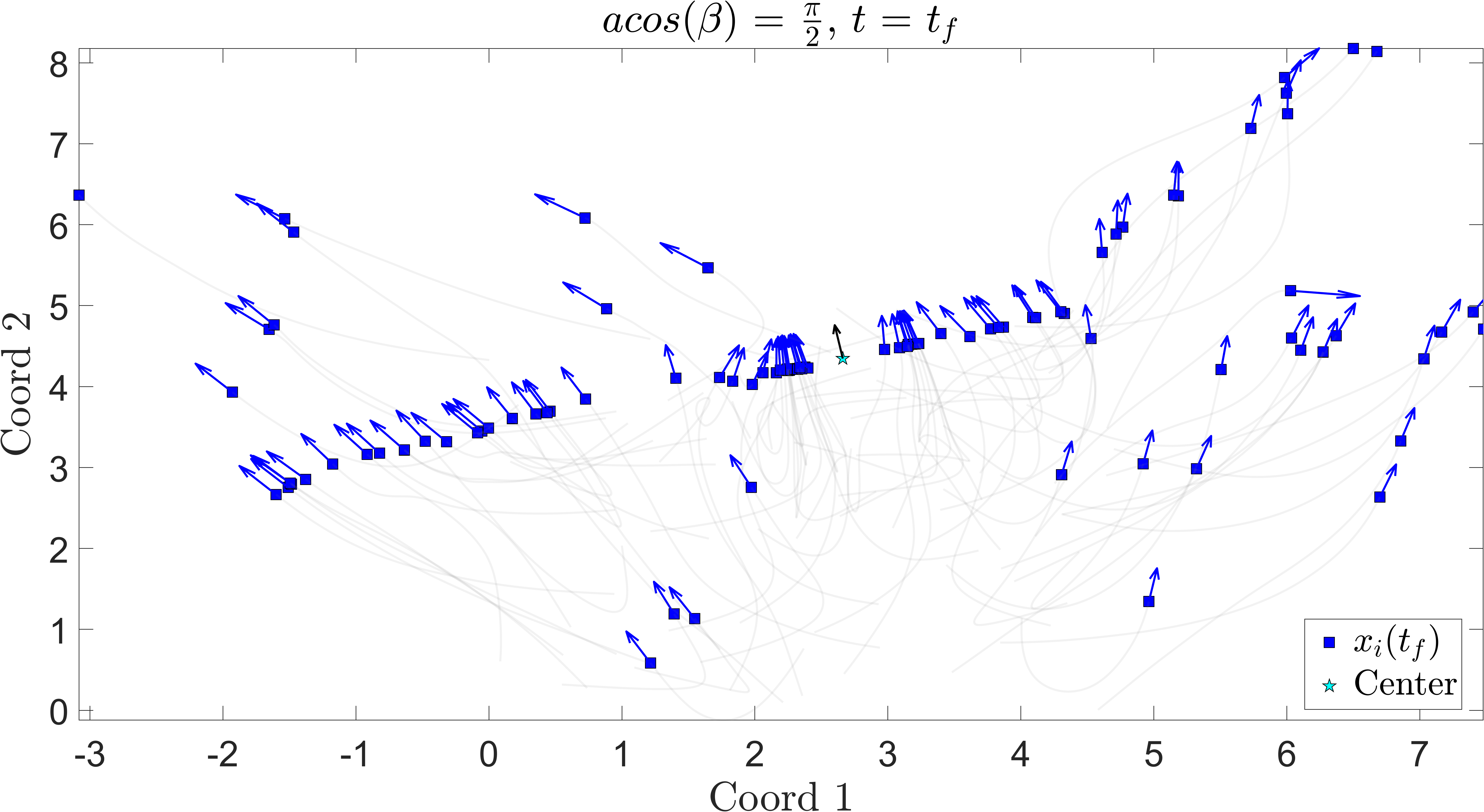}
    \caption{\rev{Trajectory plot for the second-order system with parameters taken from Table~\ref{table:2nd_order_params} with $\beta = \frac{\pi}{2}$, and initial configuration of positions/velocities is shown in Figure~\ref{fig:result_2nd_order_y0_N100}.}}
    \label{fig:result_2nd_order_yt_beta2_N100}
\end{figure}

\begin{figure}[htbp]
    \centering
    \includegraphics[width=0.48\textwidth]{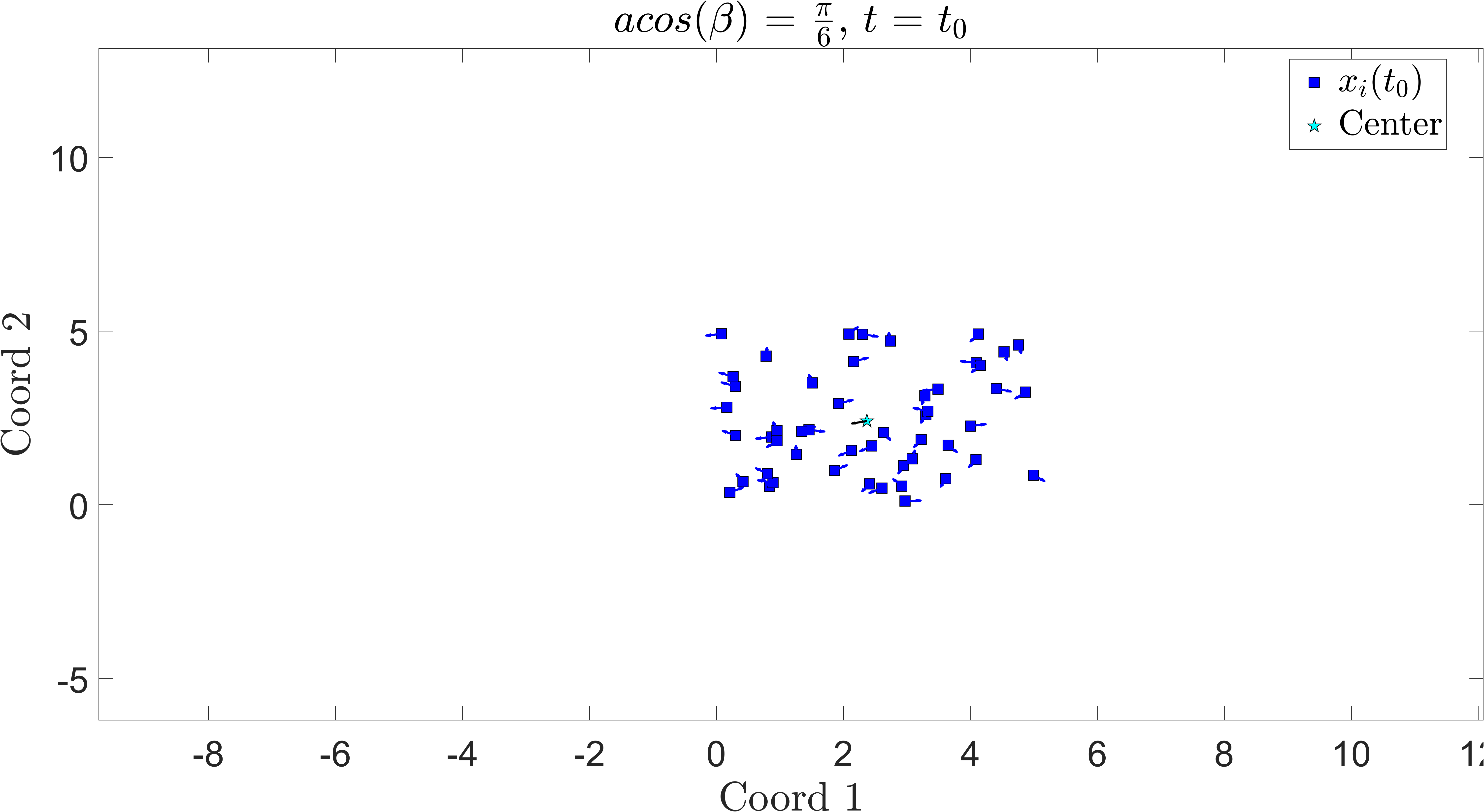}
    \caption{\rev{Initial configuration of position for second-order system with the parameters taken from Table~\ref{table:2nd_order_params} for two different $\beta$ values and $N = 50$.}}
    \label{fig:result_2nd_order_y0_N50}
\end{figure}

\begin{figure}[htbp]
    \centering
    \includegraphics[width=0.48\textwidth]{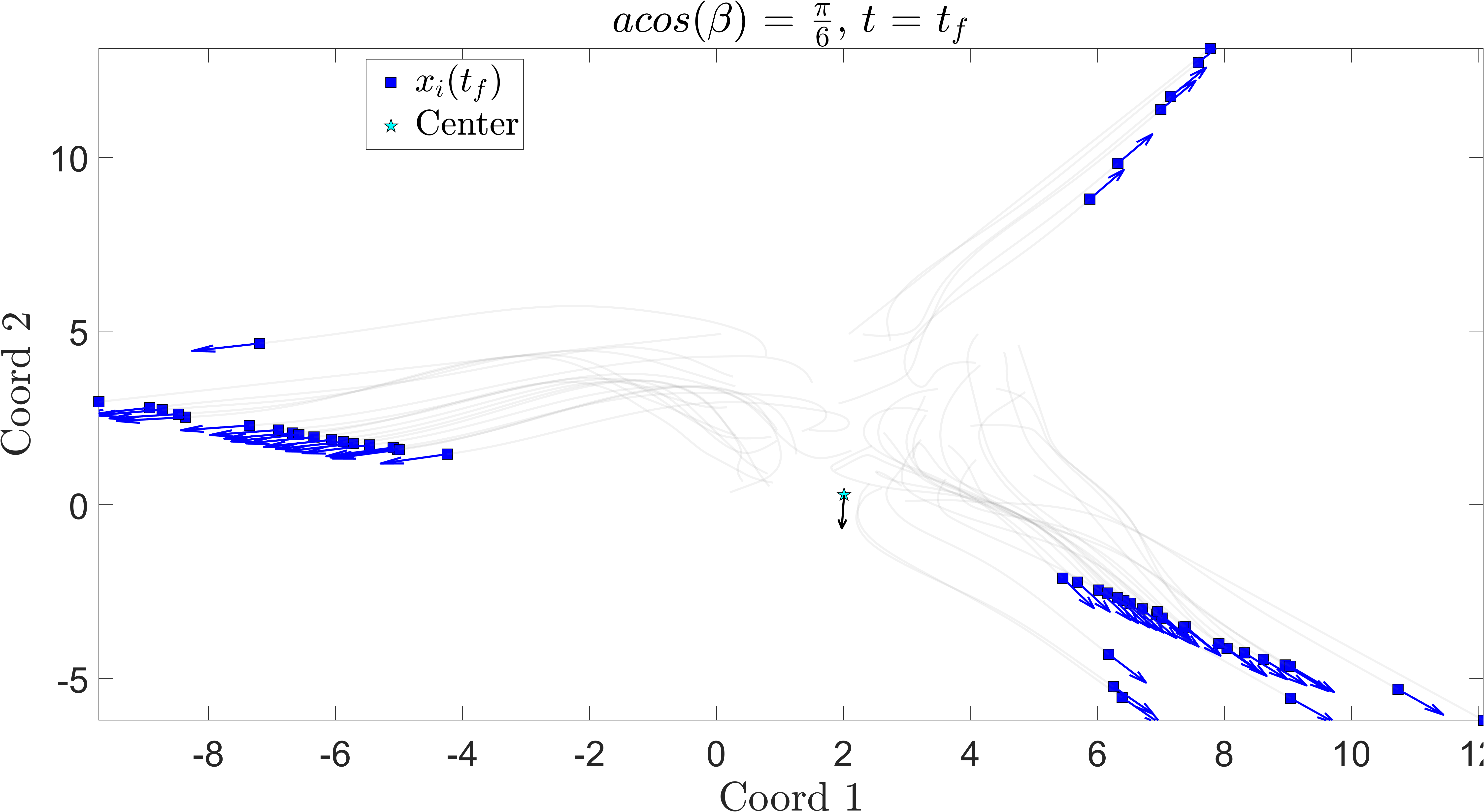}
    \caption{\rev{Trajectory plot for the second-order system with parameters taken from Table~\ref{table:2nd_order_params} with $\beta = \frac{\pi}{6}$, and initial configuration of positions/velocities is shown in Figure~\ref{fig:result_2nd_order_y0_N50}.}}
    \label{fig:result_2nd_order_yt_beta1_N50}
\end{figure}

\begin{figure}[htbp]
    \centering
    \includegraphics[width=0.48\textwidth]{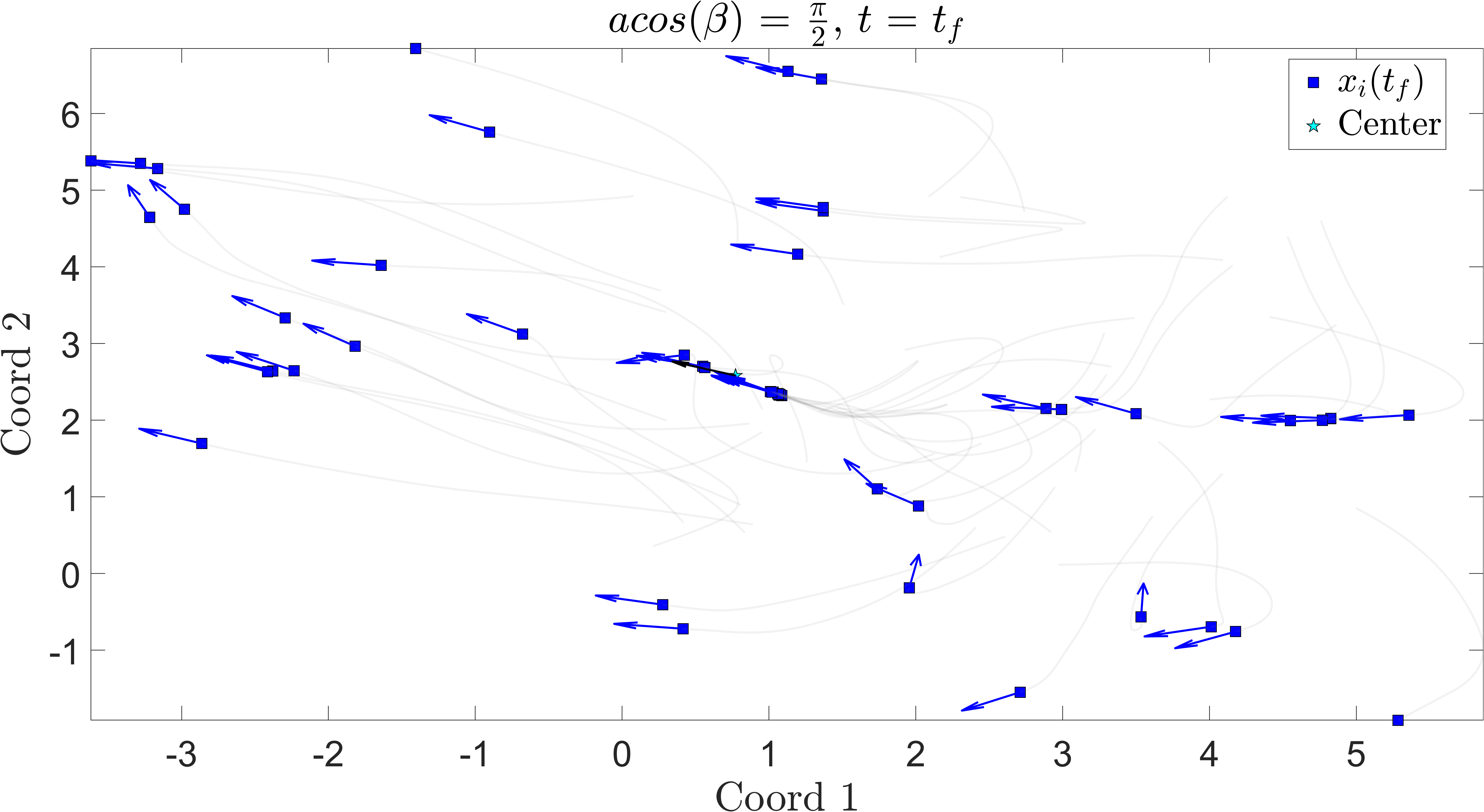}
    \caption{\rev{Trajectory plot for the second-order system with parameters taken from Table~\ref{table:2nd_order_params} with $\beta = \frac{\pi}{2}$, and initial configuration of positions/velocities is shown in Figure~\ref{fig:result_2nd_order_y0_N50}.}}
    \label{fig:result_2nd_order_yt_beta2_N50}
\end{figure}

Here $\mathbb{B}^2$ is the $2D$ unit ball \rev{centered at the origin}.  We \rev{\mysout{test}simulate} the models for different values of $\beta$ to investigate the effect of the various sizes of the neighborhood, \rev{as in Section~\ref{subsec:first_order_vary_cone} for the first-order model.  \mysout{The dynamics is evolved using MATLAB's built-in adaptive integrator $\text{ode}23$ for handling possible stiffness of the system.}}  All tests use the same initial condition, with the initial position $\bx_i(t_0)$ being an \rev{independent and identically distributed} (i.i.d) sample from $\mu^{\bx}_0$, and the initial velocity being an i.i.d sample from $\mu^{\bv}_0$.  \rev{\mysout{The $\phi$ provides attraction interaction function; where $\psi$ is the alignment interaction function, i.e. aligning the heading of the agents.}} Figure~\ref{fig:result_2nd_order_y0_N100} provides the realization of the initial configuration of positions/velocities for all the tests of different $\beta$ values \rev{investigated in this work}.  The blue dots in the figures represent the position (i.e. $\bx_i(t)$) of the agents with the yellow arrow representing the velocity (i.e. $\bv_i(t)$), whereas the \rev{\mysout{red}cyan} dot represents the center of mass position, and $\bar\bx_i(t)$ with the yellow arrow representing the center of mass velocity $\bar\bv_i(t)$.  \rev{Results are provided in Figures~\ref{fig:result_2nd_order_yt_beta1_N100}-\ref{fig:result_2nd_order_yt_beta2_N50}, where we have again considered variation in forward cone size ($\beta$) and number of agents ($N=100$ versus $N=50$).  The results are qualitatively similar to the first-order system, although the second-order system exhibits a much richer class of dynamics, as the system does not approach equilibrium configurations (this is because inter-agent forces act on accelerations and not velocities, i.e. the system is not dissipative).   Note that local groups of agents initially form curved lines of hierarchy (``leaders-followers") which eventually straighten due to the projected interaction kernel  (Figure~\ref{fig:demo_2nd_yt}).  It thus appears that the competition between line formation (induced by $\phi_1$) and flocking (induced by $\phi_2$) has made the emergent pattern more interesting.  As shown in three of the four examples of different $\beta$ (with larger $\beta$), the system as a whole can produce different flocking velocities, yet social hierarchy (in this case, lines) can emerge from the initially chaotic configuration.}

\rev{\mysout{When we switch to a smaller system, i.e. $N = 50$, we obtain the similar results (formation of geometric structures, i.e. lines, as well as flocking), shown in the following figures.}}

\noindent
\rev{\mysout{\textbf{Summary}: The competition between line formation (induced by $\phi_1$) and flocking (induced by $\phi_2$) has made the emergent pattern more interesting.  As shown in three of the four examples of different $\beta$ (with bigger $\beta$), the whole system can have different flocking velocities, yet social hierarchy (or in this case, lines) can emerge from the initially chaotic configuration. }}

\subsection{\rev{Initial conditions are not simple predictors of leader emergence}}
\label{subsec:leaders_emerge}

\rev{We note that the emergence of leaders is not a simple function of initial distance from the initial center of mass, and thus hierarchy is indeed emergent from the proposed inter-agent dynamics.  Consider Figure~\ref{fig:demo_2nd_yt}, which shows a number of flocking lines (``fingers") emerging from an initially random distribution of positions and velocities (Figure~\ref{fig:result_2nd_order_y0_N100}).  In Figure~\ref{fig:demo_2nd_yt}, we have colored red the $10$ initial agents farthest from the initial center of mass.  Note that some agents that were initially farthest from the center of mass become followers; for example, in the right-hand side, there is a line of agents with an initially distant agent (red) that becomes a ``follower."  Similarly, we see that some agents that were initially near the center of mass become ``leaders" of groups of agents.  Hence, we see that the model does indeed exhibit emergence of social hierarchy.}

\section{Discussion and conclusions}
\label{sec:discussion_conclusions}
\rev{In this work, we have presented} two collective dynamic models, namely a first-order and a second-order systems, where the emergence of social hierarchy, aka line formation, is induced from the ``looking ahead'' tendency of agents.  These two models are \rev{\mysout{simplistic}minimal} in a sense that the ``look-ahead'' tendency is implemented using a projected distance together with a forward-cone neighborhood.  These models show promising features of natural emergence of \rev{\mysout{social hierarchy, i.e. a leader/follower dynamic} geometric structures for various different kinds of initial configurations}.  More complicated patterns can be induced by using different types of communication kernels ($\phi$ or $\psi$).  \rev{We have presented numerous numerical simulations, and are currently developing a mathematical theory to rigorously understand properties of the emergence of line formation in such models. \mysout{is in the work}}.

\newpage
\bibliography{line_alignment}{}

\begin{thebibliography}{100}

\bibitem{parrish1997animal}
Julia~K Parrish and William~M Hamner.
\newblock {\em Animal groups in three dimensions: how species aggregate}.
\newblock Cambridge University Press, 1997.

\bibitem{motsch2014heterophilious}
Sebastien Motsch and Eitan Tadmor.
\newblock Heterophilious dynamics enhances consensus.
\newblock {\em SIAM review}, 56(4):577--621, 2014.

\bibitem{shvydkoy2021dynamics}
Roman Shvydkoy et~al.
\newblock {\em Dynamics and analysis of alignment models of collective
  behavior}.
\newblock Springer, 2021.

\bibitem{tadmor2021mathematics}
Eitan Tadmor.
\newblock On the mathematics of swarming: emergent behavior in alignment
  dynamics.
\newblock {\em Notices of the AMS}, 68:493--503, 2021.

\bibitem{cucker2007emergent}
Felipe Cucker and Steve Smale.
\newblock Emergent behavior in flocks.
\newblock {\em IEEE Transactions on automatic control}, 52(5):852--862, 2007.

\bibitem{cucker2007mathematics}
Felipe Cucker and Steve Smale.
\newblock On the mathematics of emergence.
\newblock {\em Japanese Journal of Mathematics}, 2(1):197--227, 2007.

\bibitem{ballerini2008}
Michele Ballerini, Nicola Cabibbo, Raphael Candelier, Andrea Cavagna, Evaristo
  Cisbani, Irene Giardina, Vivien Lecomte, Alberto Orlandi, Giorgio Parisi,
  Andrea Procaccini, et~al.
\newblock Interaction ruling animal collective behavior depends on topological
  rather than metric distance: Evidence from a field study.
\newblock {\em Proceedings of the national academy of sciences},
  105(4):1232--1237, 2008.

\bibitem{shvydkoy2020topologically}
Roman Shvydkoy and Eitan Tadmor.
\newblock Topologically based fractional diffusion and emergent dynamics with
  short-range interactions.
\newblock {\em SIAM Journal on Mathematical Analysis}, 52(6):5792--5839, 2020.

\bibitem{ha2008}
SY~Ha and E~Tadmor.
\newblock From particle to kinetic and hydrodynamic descriptions of flocking.
\newblock {\em Kinetic and Related Models}, 1(3):415--435, 2008.

\bibitem{ha2009}
S.Y. Ha and J.G. Liu.
\newblock A simple proof of the cucker-smale flocking dynamics and mean-field
  limit.
\newblock {\em Commun. Math. Sci.}, 7:297--325, 2009.

\bibitem{peszek2015discrete}
Jan Peszek.
\newblock Discrete cucker--smale flocking model with a weakly singular weight.
\newblock {\em SIAM Journal on Mathematical Analysis}, 47(5):3671--3686, 2015.

\bibitem{shvydkoy2017eulerian}
Roman Shvydkoy and Eitan Tadmor.
\newblock Eulerian dynamics with a commutator forcing.
\newblock {\em Transactions of Mathematics and its Applications}, 1(1), 2017.

\bibitem{do2018global}
Tam Do, Alexander Kiselev, Lenya Ryzhik, and Changhui Tan.
\newblock Global regularity for the fractional euler alignment system.
\newblock {\em Archive for Rational Mechanics and Analysis}, 228:1--37, 2018.

\bibitem{vicsek1995novel}
Tam{\'a}s Vicsek, Andr{\'a}s Czir{\'o}k, Eshel Ben-Jacob, Inon Cohen, and Ofer
  Shochet.
\newblock Novel type of phase transition in a system of self-driven particles.
\newblock {\em Physical review letters}, 75(6):1226, 1995.

\bibitem{shu2020flocking}
Ruiwen Shu and Eitan Tadmor.
\newblock Flocking hydrodynamics with external potentials.
\newblock {\em Archive for Rational Mechanics and Analysis}, 238:347--381,
  2020.

\bibitem{reynolds1987flocks}
Craig~W Reynolds.
\newblock Flocks, herds and schools: A distributed behavioral model.
\newblock In {\em Proceedings of the 14th annual conference on Computer
  graphics and interactive techniques}, pages 25--34, 1987.

\bibitem{vicsek2012collective}
Tam{\'a}s Vicsek and Anna Zafeiris.
\newblock Collective motion.
\newblock {\em Physics reports}, 517(3-4):71--140, 2012.

\bibitem{marchetti2013hydrodynamics}
M~Cristina Marchetti, Jean-Fran{\c{c}}ois Joanny, Sriram Ramaswamy,
  Tanniemola~B Liverpool, Jacques Prost, Madan Rao, and R~Aditi Simha.
\newblock Hydrodynamics of soft active matter.
\newblock {\em Reviews of modern physics}, 85(3):1143, 2013.

\bibitem{graner2017forms}
Fran{\c{c}}ois Graner and Daniel Riveline.
\newblock ‘the forms of tissues, or cell-aggregates’: D'arcy thompson's
  influence and its limits.
\newblock {\em Development}, 144(23):4226--4237, 2017.

\bibitem{toyoda2015cell}
Taro Toyoda, Shin-Ichi Mae, Hiromi Tanaka, Yasushi Kondo, Michinori Funato,
  Yoshiya Hosokawa, Tomomi Sudo, Yoshiya Kawaguchi, and Kenji Osafune.
\newblock Cell aggregation optimizes the differentiation of human escs and
  ipscs into pancreatic bud-like progenitor cells.
\newblock {\em Stem cell research}, 14(2):185--197, 2015.

\bibitem{zhang2010cell}
Xing Zhang, Li-hua Xu, and Qiang Yu.
\newblock Cell aggregation induces phosphorylation of pecam-1 and pyk2 and
  promotes tumor cell anchorage-independent growth.
\newblock {\em Molecular cancer}, 9(1):1--11, 2010.

\bibitem{bayoussef2012aggregation}
Zahia Bayoussef, James~E Dixon, Snjezana Stolnik, and Kevin~M Shakesheff.
\newblock Aggregation promotes cell viability, proliferation, and
  differentiation in an in vitro model of injection cell therapy.
\newblock {\em Journal of tissue engineering and regenerative medicine},
  6(10):e61--e73, 2012.

\bibitem{glinel2012antibacterial}
Karine Glinel, Pascal Thebault, Vincent Humblot, Claire-Marie Pradier, and
  Thierry Jouenne.
\newblock Antibacterial surfaces developed from bio-inspired approaches.
\newblock {\em Acta biomaterialia}, 8(5):1670--1684, 2012.

\bibitem{green1999adhesion}
Shane~K Green, Andrea Frankel, and Robert~S Kerbel.
\newblock Adhesion-dependent multicellular drug resistance.
\newblock {\em Anti-cancer drug design}, 14(2):153--168, 1999.

\bibitem{croix1997cell}
Brad~St Croix and Robert~S Kerbel.
\newblock Cell adhesion and drug resistance in cancer.
\newblock {\em Current opinion in oncology}, 9(6):549--556, 1997.

\bibitem{brown2016aggregation}
Joel~S Brown, Jessica~J Cunningham, and Robert~A Gatenby.
\newblock Aggregation effects and population-based dynamics as a source of
  therapy resistance in cancer.
\newblock {\em IEEE Transactions on Biomedical Engineering}, 64(3):512--518,
  2016.

\bibitem{lavi2013role}
Orit Lavi, James~M Greene, Doron Levy, and Michael~M Gottesman.
\newblock The role of cell density and intratumoral heterogeneity in multidrug
  resistancemodeling the intratumoral heterogeneity in multidrug resistance.
\newblock {\em Cancer research}, 73(24):7168--7175, 2013.

\bibitem{friedl2009collective}
Peter Friedl and Darren Gilmour.
\newblock Collective cell migration in morphogenesis, regeneration and cancer.
\newblock {\em Nature reviews Molecular cell biology}, 10(7):445--457, 2009.

\bibitem{theveneau2012neural}
Eric Theveneau and Roberto Mayor.
\newblock Neural crest migration: interplay between chemorepellents,
  chemoattractants, contact inhibition, epithelial--mesenchymal transition, and
  collective cell migration.
\newblock {\em Wiley Interdisciplinary Reviews: Developmental Biology},
  1(3):435--445, 2012.

\bibitem{varuni2017phototaxis}
P~Varuni, Shakti~N Menon, and Gautam~I Menon.
\newblock Phototaxis as a collective phenomenon in cyanobacterial colonies.
\newblock {\em Scientific reports}, 7(1):1--10, 2017.

\bibitem{morrell2008mechanisms}
Lesley~J Morrell and Richard James.
\newblock Mechanisms for aggregation in animals: rule success depends on
  ecological variables.
\newblock {\em Behavioral Ecology}, 19(1):193--201, 2008.

\bibitem{couzin2002collective}
Iain~D Couzin, Jens Krause, Richard James, Graeme~D Ruxton, and Nigel~R Franks.
\newblock Collective memory and spatial sorting in animal groups.
\newblock {\em Journal of theoretical biology}, 218(1):1--11, 2002.

\bibitem{couzin2003self}
Iain~D Couzin and Nigel~R Franks.
\newblock Self-organized lane formation and optimized traffic flow in army
  ants.
\newblock {\em Proceedings of the Royal Society of London. Series B: Biological
  Sciences}, 270(1511):139--146, 2003.

\bibitem{motsch2011}
Sebastien Motsch and Eitan Tadmor.
\newblock A new model for self-organized dynamics and its flocking behavior.
\newblock {\em Journal of Statistical Physics}, 144(5):923--947, 2011.

\bibitem{conradt2005consensus}
Larissa Conradt and Timothy~J Roper.
\newblock Consensus decision making in animals.
\newblock {\em Trends in ecology \& evolution}, 20(8):449--456, 2005.

\bibitem{krause2000fish}
Jens Krause, Daniel~J Hoare, Darren Croft, James Lawrence, Ashley Ward,
  Graeme~D Ruxton, Jean-Guy~J Godin, and Richard James.
\newblock Fish shoal composition: mechanisms and constraints.
\newblock {\em Proceedings of the Royal Society of London. Series B: Biological
  Sciences}, 267(1456):2011--2017, 2000.

\bibitem{hemelrijk2015increased}
CK~Hemelrijk, DAP Reid, H~Hildenbrandt, and JT~Padding.
\newblock The increased efficiency of fish swimming in a school.
\newblock {\em Fish and Fisheries}, 16(3):511--521, 2015.

\bibitem{marras2015fish}
Stefano Marras, Shaun~S Killen, Jan Lindstr{\"o}m, David~J McKenzie, John~F
  Steffensen, and Paolo Domenici.
\newblock Fish swimming in schools save energy regardless of their spatial
  position.
\newblock {\em Behavioral ecology and sociobiology}, 69(2):219--226, 2015.

\bibitem{cavagna2008starflag}
Andrea Cavagna, Irene Giardina, Alberto Orlandi, Giorgio Parisi, Andrea
  Procaccini, Massimiliano Viale, and Vladimir Zdravkovic.
\newblock The starflag handbook on collective animal behaviour: Part i,
  empirical methods.
\newblock {\em arXiv preprint arXiv:0802.1668}, 2008.

\bibitem{ballerini2008empirical}
Michele Ballerini, Nicola Cabibbo, Raphael Candelier, Andrea Cavagna, Evaristo
  Cisbani, Irene Giardina, Alberto Orlandi, Giorgio Parisi, Andrea Procaccini,
  Massimiliano Viale, et~al.
\newblock Empirical investigation of starling flocks: a benchmark study in
  collective animal behaviour.
\newblock {\em Animal behaviour}, 76(1):201--215, 2008.

\bibitem{bajec2009organized}
Iztok~Lebar Bajec and Frank~H Heppner.
\newblock Organized flight in birds.
\newblock {\em Animal Behaviour}, 78(4):777--789, 2009.

\bibitem{ling2019local}
Hangjian Ling, Guillam~E Mclvor, Kasper van~der Vaart, Richard~T Vaughan, Alex
  Thornton, and Nicholas~T Ouellette.
\newblock Local interactions and their group-level consequences in flocking
  jackdaws.
\newblock {\em Proceedings of the Royal Society B}, 286(1906):20190865, 2019.

\bibitem{hughey2018challenges}
Lacey~F Hughey, Andrew~M Hein, Ariana Strandburg-Peshkin, and Frants~H Jensen.
\newblock Challenges and solutions for studying collective animal behaviour in
  the wild.
\newblock {\em Philosophical Transactions of the Royal Society B: Biological
  Sciences}, 373(1746):20170005, 2018.

\bibitem{goodenough2017birds}
Anne~E Goodenough, Natasha Little, William~S Carpenter, and Adam~G Hart.
\newblock Birds of a feather flock together: Insights into starling murmuration
  behaviour revealed using citizen science.
\newblock {\em PloS one}, 12(6):e0179277, 2017.

\bibitem{mueller2013social}
Thomas Mueller, Robert~B O’Hara, Sarah~J Converse, Richard~P Urbanek, and
  William~F Fagan.
\newblock Social learning of migratory performance.
\newblock {\em Science}, 341(6149):999--1002, 2013.

\bibitem{riters2019birds}
Lauren~V Riters, Cynthia~A Kelm-Nelson, and Jeremy~A Spool.
\newblock Why do birds flock? a role for opioids in the reinforcement of
  gregarious social interactions.
\newblock {\em Frontiers in Physiology}, 10:421, 2019.

\bibitem{sarfati2021self}
Rapha{\"e}l Sarfati, Julie~C Hayes, and Orit Peleg.
\newblock Self-organization in natural swarms of photinus carolinus synchronous
  fireflies.
\newblock {\em Science Advances}, 7(28):eabg9259, 2021.

\bibitem{buck1988synchronous}
John Buck.
\newblock Synchronous rhythmic flashing of fireflies. ii.
\newblock {\em The Quarterly review of biology}, 63(3):265--289, 1988.

\bibitem{penn2016network}
Yaron Penn, Menahem Segal, and Elisha Moses.
\newblock Network synchronization in hippocampal neurons.
\newblock {\em Proceedings of the National Academy of Sciences},
  113(12):3341--3346, 2016.

\bibitem{ben2005opinion}
Eli Ben-Naim.
\newblock Opinion dynamics: rise and fall of political parties.
\newblock {\em EPL (Europhysics Letters)}, 69(5):671, 2005.

\bibitem{vicsek2001question}
Tamas Vicsek.
\newblock A question of scale.
\newblock {\em Nature}, 411(6836):421--421, 2001.

\bibitem{delgado2007use}
Carlos Delgado-Mata, Jesus~Ibanez Martinez, Simon Bee, Rocio Ruiz-Rodarte, and
  Ruth Aylett.
\newblock On the use of virtual animals with artificial fear in virtual
  environments.
\newblock {\em New Generation Computing}, 25(2):145--169, 2007.

\bibitem{braga2018collision}
Rafael~G Braga, Roberto~C Da~Silva, Alexandre~CB Ramos, and Felix Mora-Camino.
\newblock Collision avoidance based on reynolds rules: A case study using
  quadrotors.
\newblock In {\em Information Technology-New Generations: 14th International
  Conference on Information Technology}, pages 773--780. Springer, 2018.

\bibitem{kennedy1995particle}
James Kennedy and Russell Eberhart.
\newblock Particle swarm optimization.
\newblock In {\em Proceedings of ICNN'95-international conference on neural
  networks}, volume~4, pages 1942--1948. IEEE, 1995.

\bibitem{hartman2006autonomous}
Christopher Hartman and Bedrich Benes.
\newblock Autonomous boids.
\newblock {\em Computer Animation and Virtual Worlds}, 17(3-4):199--206, 2006.

\bibitem{shu2021anticipation}
Ruiwen Shu and Eitan Tadmor.
\newblock Anticipation breeds alignment.
\newblock {\em Archive for Rational Mechanics and Analysis}, 240:203--241,
  2021.

\bibitem{carrillo2008double}
Jos{\'e}~A Carrillo, Maria~R D'Orsogna, and Vladislav Panferov.
\newblock Double milling in self-propelled swarms from kinetic theory.
\newblock {\em Kinetic and Related Models}, 2:363--378, 2009.

\bibitem{ariel2015locust}
Gil Ariel and Amir Ayali.
\newblock Locust collective motion and its modeling.
\newblock {\em PLOS Computational Biology}, 11(12):e1004522, 2015.

\bibitem{buhl2006disorder}
Jerome Buhl, David~JT Sumpter, Iain~D Couzin, Joe~J Hale, Emma Despland,
  Edgar~R Miller, and Steve~J Simpson.
\newblock From disorder to order in marching locusts.
\newblock {\em Science}, 312(5778):1402--1406, 2006.

\bibitem{wilson2004basking}
Steven~G Wilson.
\newblock Basking sharks (cetorhinus maximus) schooling in the southern gulf of
  maine.
\newblock {\em Fisheries Oceanography}, 13(4):283--286, 2004.

\bibitem{newRef01}
Malet-Engra Gema, Yu~Weimiao, Oldani Admanda, Rey-Barroso Javier, Gov~Nir S.,
  Scita Giorgio, and Dupr\'{e} Lo\"{i}c.
\newblock Collective cell motility promotes chemotactic prowess and resistance
  to chemorepulsion.
\newblock {\em Curr Biol.}, 25(2):242 -- 250, 2015.

\bibitem{newRef02}
Katherine Copenhagen, Gema Malet-Engra, Weimiao Yu, Giorgio Scita, Nir Gov, and
  Ajay Gopinathan.
\newblock Frustration-induced phases in migrating cell clusters.
\newblock {\em Science Advances}, 4(9):eaar8483, 2018.

\bibitem{schneirla1944unique}
Theodore~Christian Schneirla et~al.
\newblock A unique case of circular milling in ants, considered in relation to
  trail following and the general problem of orientation.
\newblock {\em American Museum Novitates}, 1253, 1944.

\bibitem{antoniou2013congestion}
Pavlos Antoniou, Andreas Pitsillides, Tim Blackwell, Andries Engelbrecht, and
  Loizos Michael.
\newblock Congestion control in wireless sensor networks based on bird flocking
  behavior.
\newblock {\em Computer Networks}, 57(5):1167--1191, 2013.

\bibitem{hajihassani2018applications}
M~Hajihassani, D~Jahed~Armaghani, and R~Kalatehjari.
\newblock Applications of particle swarm optimization in geotechnical
  engineering: a comprehensive review.
\newblock {\em Geotechnical and Geological Engineering}, 36(2):705--722, 2018.

\bibitem{vasarhelyi2018optimized}
G{\'a}bor V{\'a}s{\'a}rhelyi, Csaba Vir{\'a}gh, Gerg{\H{o}} Somorjai, Tam{\'a}s
  Nepusz, Agoston~E Eiben, and Tam{\'a}s Vicsek.
\newblock Optimized flocking of autonomous drones in confined environments.
\newblock {\em Science Robotics}, 3(20):eaat3536, 2018.

\bibitem{cao2012overview}
Yongcan Cao, Wenwu Yu, Wei Ren, and Guanrong Chen.
\newblock An overview of recent progress in the study of distributed
  multi-agent coordination.
\newblock {\em IEEE Transactions on Industrial informatics}, 9(1):427--438,
  2012.

\bibitem{olfati2006flocking}
Reza Olfati-Saber.
\newblock Flocking for multi-agent dynamic systems: Algorithms and theory.
\newblock {\em IEEE Transactions on automatic control}, 51(3):401--420, 2006.

\bibitem{couzin2009collective}
Iain~D Couzin.
\newblock Collective cognition in animal groups.
\newblock {\em Trends in cognitive sciences}, 13(1):36--43, 2009.

\bibitem{vedel2013migration}
S{\o}ren Vedel, Sava{\c{s}} Tay, Darius~M Johnston, Henrik Bruus, and Stephen~R
  Quake.
\newblock Migration of cells in a social context.
\newblock {\em Proceedings of the National Academy of Sciences},
  110(1):129--134, 2013.

\bibitem{qin2021roles}
Lei Qin, Dazhi Yang, Weihong Yi, Huiling Cao, and Guozhi Xiao.
\newblock Roles of leader and follower cells in collective cell migration.
\newblock {\em Molecular biology of the cell}, 32(14):1267--1272, 2021.

\bibitem{brian2012social}
Michael~Vaughan Brian.
\newblock {\em Social insects: ecology and behavioural biology}.
\newblock Springer Science \& Business Media, 2012.

\bibitem{clutton2016mammal}
Tim Clutton-Brock.
\newblock {\em Mammal societies}.
\newblock John Wiley \& Sons, 2016.

\bibitem{alexandre2015chemotaxis}
Gladys Alexandre.
\newblock Chemotaxis control of transient cell aggregation.
\newblock {\em Journal of bacteriology}, 197(20):3230--3237, 2015.

\bibitem{chau2017emergent}
Rosanna Man~Wah Chau, Devaki Bhaya, and Kerwyn~Casey Huang.
\newblock Emergent phototactic responses of cyanobacteria under complex light
  regimes.
\newblock {\em MBio}, 8(2):e02330--16, 2017.

\bibitem{ursell2013motility}
Tristan Ursell, Rosanna Man~Wah Chau, Susanne Wisen, Devaki Bhaya, and
  Kerwyn~Casey Huang.
\newblock Motility enhancement through surface modification is sufficient for
  cyanobacterial community organization during phototaxis.
\newblock {\em PLoS computational biology}, 9(9):e1003205, 2013.

\bibitem{poujade2007collective}
Mathieu Poujade, Erwan Grasland-Mongrain, A~Hertzog, J~Jouanneau, Philippe
  Chavrier, Beno{\^\i}t Ladoux, Axel Buguin, and Pascal Silberzan.
\newblock Collective migration of an epithelial monolayer in response to a
  model wound.
\newblock {\em Proceedings of the National Academy of Sciences},
  104(41):15988--15993, 2007.

\bibitem{gov2007collective}
Nir~S Gov.
\newblock Collective cell migration patterns: follow the leader.
\newblock {\em Proceedings of the National Academy of Sciences},
  104(41):15970--15971, 2007.

\bibitem{ladoux2017mechanobiology}
Ladoux Benoit and M{\'e}ge Ren{\'e}-Marc.
\newblock Mechanobiology of collective cell behaviours.
\newblock {\em Nature reviews Molecular cell biology}, 18(12):743--757, 2017.

\bibitem{sengupta2021principles}
Shuvasree SenGupta, Carole~A Parent, and James~E Bear.
\newblock The principles of directed cell migration.
\newblock {\em Nature Reviews Molecular Cell Biology}, 22(8):529--547, 2021.

\bibitem{czaczkes2015trail}
Tomer~J Czaczkes, Christoph Gr{\"u}ter, and Francis~LW Ratnieks.
\newblock Trail pheromones: an integrative view of their role in social insect
  colony organization.
\newblock {\em Annual review of entomology}, 60:581--599, 2015.

\bibitem{perna2012individual}
Andrea Perna, Boris Granovskiy, Simon Garnier, Stamatios~C Nicolis, Marjorie
  Lab{\'e}dan, Guy Theraulaz, Vincent Fourcassi{\'e}, and David~JT Sumpter.
\newblock Individual rules for trail pattern formation in argentine ants
  (linepithema humile).
\newblock {\em PLoS computational biology}, 8(7):e1002592, 2012.

\bibitem{gordon2019ecology}
Deborah~M Gordon et~al.
\newblock The ecology of collective behavior in ants.
\newblock {\em Annu. Rev. Entomol}, 64:35--50, 2019.

\bibitem{feinerman2018physics}
Ofer Feinerman, Itai Pinkoviezky, Aviram Gelblum, Ehud Fonio, and Nir~S Gov.
\newblock The physics of cooperative transport in groups of ants.
\newblock {\em Nature Physics}, 14(7):683--693, 2018.

\bibitem{theveneau2017leaders}
Eric Theveneau and Claudia Linker.
\newblock Leaders in collective migration: are front cells really endowed with
  a particular set of skills?
\newblock {\em F1000Research}, 6, 2017.

\bibitem{kozyrska2022p53}
Kasia Kozyrska, Giulia Pilia, Medhavi Vishwakarma, Laura Wagstaff, Maja
  Goschorska, Silvia Cirillo, Saad Mohamad, Kelli Gallacher, Rafael~E
  Carazo~Salas, and Eugenia Piddini.
\newblock p53 directs leader cell behavior, migration, and clearance during
  epithelial repair.
\newblock {\em Science}, 375(6581):eabl8876, 2022.

\bibitem{bullo2019lectures}
Francesco Bullo.
\newblock {\em Lectures on network systems}, volume~1.
\newblock Kindle Direct Publishing Santa Barbara, CA, 2019.

\bibitem{levine2000self}
Herbert Levine, Wouter-Jan Rappel, and Inon Cohen.
\newblock Self-organization in systems of self-propelled particles.
\newblock {\em Physical Review E}, 63(1):017101, 2000.

\bibitem{d2006self}
Maria~R D’Orsogna, Yao-Li Chuang, Andrea~L Bertozzi, and Lincoln~S Chayes.
\newblock Self-propelled particles with soft-core interactions: patterns,
  stability, and collapse.
\newblock {\em Physical review letters}, 96(10):104302, 2006.

\bibitem{chate2008collective}
Hugues Chat{\'e}, Francesco Ginelli, Guillaume Gr{\'e}goire, and Franck
  Raynaud.
\newblock Collective motion of self-propelled particles interacting without
  cohesion.
\newblock {\em Physical Review E}, 77(4):046113, 2008.

\bibitem{chuang2007state}
Yao-Li Chuang, Maria~R D’orsogna, Daniel Marthaler, Andrea~L Bertozzi, and
  Lincoln~S Chayes.
\newblock State transitions and the continuum limit for a 2d interacting,
  self-propelled particle system.
\newblock {\em Physica D: Nonlinear Phenomena}, 232(1):33--47, 2007.

\bibitem{minakowski2019}
Piotr Minakowski, Piotr~B Mucha, Jan Peszek, and Ewelina Zatorska.
\newblock Singular cucker--smale dynamics.
\newblock In {\em Active Particles, Volume 2}, pages 201--243. Springer, 2019.

\bibitem{choi2019}
Young-Pil Choi, Dante Kalise, Jan Peszek, and Andr{\'e}s~A Peters.
\newblock A collisionless singular cucker--smale model with decentralized
  formation control.
\newblock {\em SIAM Journal on Applied Dynamical Systems}, 18(4):1954--1981,
  2019.

\bibitem{carrillo2010asymptotic}
Jos{\'e}~A Carrillo, Massimo Fornasier, Jes{\'u}s Rosado, and Giuseppe Toscani.
\newblock Asymptotic flocking dynamics for the kinetic cucker--smale model.
\newblock {\em SIAM Journal on Mathematical Analysis}, 42(1):218--236, 2010.

\bibitem{carrillo2010particle}
Jos{\'e}~A Carrillo, Massimo Fornasier, Giuseppe Toscani, and Francesco Vecil.
\newblock Particle, kinetic, and hydrodynamic models of swarming.
\newblock {\em Mathematical modeling of collective behavior in socio-economic
  and life sciences}, pages 297--336, 2010.

\bibitem{carrillo2014derivation}
Jos{\'e}~Antonio Carrillo, Young-Pil Choi, and Maxime Hauray.
\newblock The derivation of swarming models: mean-field limit and wasserstein
  distances.
\newblock {\em Collective dynamics from bacteria to crowds}, 553:1--46, 2014.

\bibitem{choi2017emergent}
Young-Pil Choi, Seung-Yeal Ha, and Zhuchun Li.
\newblock Emergent dynamics of the cucker--smale flocking model and its
  variants.
\newblock {\em Active Particles, Volume 1: Advances in Theory, Models, and
  Applications}, pages 299--331, 2017.

\bibitem{park2010cucker}
Jaemann Park, H~Jin Kim, and Seung-Yeal Ha.
\newblock Cucker-smale flocking with inter-particle bonding forces.
\newblock {\em IEEE Transactions on Automatic Control}, 55(11):2617--2623,
  2010.

\bibitem{djokam2022generalized}
Guy~A Djokam and Muruhan Rathinam.
\newblock A generalized model of flocking with steering.
\newblock {\em SIAM Journal on Applied Dynamical Systems}, 21(2):1352--1381,
  2022.

\bibitem{ha2009stochastic}
K.~Ha~S.Y., Lee and Levy D.
\newblock A simple proof of the cucker-smale flocking dynamics and mean-field
  limit.
\newblock {\em Commun. Math. Sci.}, 7(2):453--469, 2009.

\bibitem{galante2011stochastic}
Amanda Galante, Susanne Wisen, Devaki Bhaya, and Doron Levy.
\newblock Stochastic models and simulations of phototaxis.
\newblock {\em Unifying Themes in Complex Systems}, 8:105--119, 2011.

\bibitem{ha2009particle}
Seung-Yeal Ha and Doron Levy.
\newblock Particle, kinetic and fluid models for phototaxis.
\newblock {\em Discrete Contin. Dyn. Syst. Ser. B}, 12(1):77--108, 2009.

\bibitem{bhaya2008group}
Devaki Bhaya, Doron Levy, and Tiago Requeijo.
\newblock Group dynamics of phototaxis: Interacting stochastic many-particle
  systems and their continuum limit.
\newblock In {\em Hyperbolic Problems: Theory, Numerics, Applications}, pages
  145--159. Springer, 2008.

\bibitem{levy2008modeling}
Doron Levy and Tiago Requeijo.
\newblock Modeling group dynamics of phototaxis: from particle systems to pdes.
\newblock {\em Discrete and Continuous Dynamical Systems Series B}, 9(1):103,
  2008.

\bibitem{levy2008stochastic}
Doron Levy and Tiago Requeijo.
\newblock Stochastic models for phototaxis.
\newblock {\em Bulletin of Mathematical Biology}, 70:1684--1706, 2008.

\bibitem{risser2013comparative}
Douglas~D Risser and John~C Meeks.
\newblock Comparative transcriptomics with a motility-deficient mutant leads to
  identification of a novel polysaccharide secretion system in n ostoc
  punctiforme.
\newblock {\em Molecular microbiology}, 87(4):884--893, 2013.

\bibitem{menon2020information}
Shakti~N Menon, P~Varuni, and Gautam~I Menon.
\newblock Information integration and collective motility in phototactic
  cyanobacteria.
\newblock {\em PLoS Computational Biology}, 16(4):e1007807, 2020.

\bibitem{amar2016collective}
M~Ben Amar.
\newblock Collective chemotaxis and segregation of active bacterial colonies.
\newblock {\em Scientific Reports}, 6(1):21269, 2016.

\bibitem{keller1971model}
Evelyn~F Keller and Lee~A Segel.
\newblock Model for chemotaxis.
\newblock {\em Journal of theoretical biology}, 30(2):225--234, 1971.

\bibitem{keller1971traveling}
Evelyn~F Keller and Lee~A Segel.
\newblock Traveling bands of chemotactic bacteria: a theoretical analysis.
\newblock {\em Journal of theoretical biology}, 30(2):235--248, 1971.

\bibitem{alert2022cellular}
Ricard Alert, Alejandro Mart{\'\i}nez-Calvo, and Sujit~S Datta.
\newblock Cellular sensing governs the stability of chemotactic fronts.
\newblock {\em Physical review letters}, 128(14):148101, 2022.

\bibitem{shen2008cucker}
Jackie Shen.
\newblock Cucker--smale flocking under hierarchical leadership.
\newblock {\em SIAM Journal on Applied Mathematics}, 68(3):694--719, 2008.

\bibitem{shao2018leader}
Jinliang Shao, Wei~Xing Zheng, Ting-Zhu Huang, and Adrian~N Bishop.
\newblock On leader--follower consensus with switching topologies: An analysis
  inspired by pigeon hierarchies.
\newblock {\em IEEE Transactions on Automatic Control}, 63(10):3588--3593,
  2018.

\bibitem{aureli2010coordination}
Matteo Aureli and Maurizio Porfiri.
\newblock Coordination of self-propelled particles through external leadership.
\newblock {\em Europhysics Letters}, 92(4):40004, 2010.

\bibitem{newRef03}
Mark Shirley, Shlomovitz Roie, Gov~Nir S., Roujade Mathieu, Grasland-Mongrain
  Erwan, and Silberzan Pascal.
\newblock Physical model of the dynamic instability in an expanding cell
  culture.
\newblock {\em Biophys J.}, 98(3):361 -- 370, 2010.

\bibitem{newRef04}
Tarle Victoria, Ravasio Andrea, Hakim Vincent, and Gov~Nir S.
\newblock Modeling the finger instability in an expanding cell monolayer.
\newblock {\em Integr Biol (Camb)}, 7(10):1218 -- 1227, 2015.

\bibitem{newRef05}
Victoria Tarle, Estelle Gauquelin, Sri Ram~Krishna Vedula, Joseph
  d’'Alessandro, Chwee~Teck Lim, Beno\^{i}t Ladoux, and Nir~S. Gov.
\newblock Modeling collective cell migration in geometric confinement.
\newblock {\em Physical Biology}, 14, 2017.

\bibitem{schuergers2016cyanobacteria}
Nils Schuergers, Tchern Lenn, Ronald Kampmann, Markus~V Meissner, Tiago
  Esteves, Maja Temerinac-Ott, Jan~G Korvink, Alan~R Lowe, Conrad~W Mullineaux,
  and Annegret Wilde.
\newblock Cyanobacteria use micro-optics to sense light direction.
\newblock {\em Elife}, 5:e12620, 2016.

\bibitem{chernetsov2017migratory}
Nikita Chernetsov, Alexander Pakhomov, Dmitry Kobylkov, Dmitry Kishkinev,
  Richard~A Holland, and Henrik Mouritsen.
\newblock Migratory eurasian reed warblers can use magnetic declination to
  solve the longitude problem.
\newblock {\em Current Biology}, 27(17):2647--2651, 2017.

\end{thebibliography}
\bibliographystyle{unsrt}
\end{document}